%% file: stpa.tex
\let\oldvec\vec
\documentclass[final]{llncs}
\let\vec\oldvec
\usepackage[utf8]{inputenc}
\usepackage{amsmath}
\usepackage{amssymb}
\usepackage{cite}
\usepackage[obeyFinal]{todonotes}
\usepackage{xspace}
\usepackage{tikz}
\usepackage{mathpartir}
\usepackage{stmaryrd}
\usepackage{datetime}
\usepackage{ifdraft}
\usepackage{ifthen}
\usepackage{etoolbox}
\usepackage{listings}
\usepackage{url}
\usepackage{mathtools}
\usepackage[capitalize]{cleveref}

\usepackage{newunicodechar}
\newunicodechar{µ}{$\mu$}
\newunicodechar{∀}{$\forall$}
\newunicodechar{α}{$\alpha$}
\newunicodechar{∉}{$\notin$}
\newunicodechar{∧}{$\wedge$}
\newunicodechar{∈}{$\in$}
\newunicodechar{∨}{$\vee$}

\include{commonmacros}

\include{localmacros}

\crefname{equation}{eq.}{eqs.}

\newboolean{IsExtVers}
\booltrue{IsExtVers}

\urldef{\maila}\path|{jakobro,thiemann}@informatik.uni-freiburg.de|

\title{A Falsification View of Success Typing}
\subtitle{\ifIsExtVers (extended version)\\\else\fi\ifoptionfinal{}{\textcolor{red}{--- DRAFT \today\ \currenttime\ ---}}}

\author{Robert Jakob \and Peter Thiemann}
\institute{%
  University of Freiburg, Germany\\
  \maila
}

\begin{document}

\maketitle

\input{abstract}

\section{Introduction}
\label{sec:Introduction}

Dynamic languages like JavaScript, Python, and Erlang are increasingly used in
application domains where reliability and robustness matters. Their advantages
lie in the provision of domain specific libraries, flexibility, and
expressiveness, which enables rapid prototyping. However, massive unit testing
with all its drawbacks is the primary method of discovering errors:
static analysis is often not applicable because it either yields many
false positives or restricts the expressiveness. Verification is
feasible but cumbersome (see for example the JavaScript formalization
effort \cite{GardnerMaffeisSmith2012,BodinChargueraudFilarettiGardnerMaffeisNaudziunieneSchmittSmith2014}). Moreover,
it requires a major effort. 

Unit testing with good code coverage is not straightforward to
achieve, either. As the development of meaningful unit tests is also 
cumbersome and time consuming, the lack of static analyses that permit error
detection prior to execution is one of the major drawbacks of dynamic
languages.

Classical static analyses and type systems guarantee the absence of a
particular class of errors: the program cannot \emph{go wrong}. Imposing such
a system on a dynamic language deprives it of its major attraction for certain
programmers: the ability to write code without being restricted by a formal
framework. Even suggesting such a framework would come close to treason.
Furthermore,  programmers are confused by false positives or error messages
they do not
understand~\cite{BesseyBlockChelfChouFultonHallemGrosKamskyMcPeakEngler2010}.
However, an analysis that only reports problems that would definitely lead to
an error during execution could be acceptable. This point of view leads to the
idea of a success typing.

In a standard type system, the typing $F: \tau_1\to\tau_2$ means that an
application of $F$ to an argument $v$ of type $\tau_1$ yields a result of type
$\tau_2$ if $F (v)$ terminates normally. If type checking for the
system is decidable, then there are programs which do not
lead to type mismatches when executed, but which are rejected by the
type system.  A trivial 
example is a conditional that returns values of different types in its
branches, but semantically it is clear that only the first branch can
ever be executed.

In contrast, a success type system guarantees that for all arguments
$v$ \textbf{not} of type $\tau_1$, the function application $F(v)$
leads to a run-time error (or nontermination).  For an argument $v$ of
type $\tau_1$, 
success typing gives the same guarantees as traditional typing:
$F(v)\in\tau_2$ if it terminates normally.  By necessity, the
guarantee of the run-time error is also an approximation, but success
typing must approximate in the other direction as a standard type
system. Hence, the ``standard part'' of a success type usually gives a
weaker guarantee than a standard type.
In model checking terms, a standard type system performs
verification whereas success typing seems related to
falsification~\cite{BallKupfermannYorsh2005}: its goal is the
detection of errors rather than proving the absence of them.

\subsection{Success Typings in Erlang}

Erlang is a dynamically typed functional programming language with
commercial uses in e-commerce, telephony, and instant messaging.  Besides
the usual numeric and string types, Erlang includes an atom data type for
symbols and tuples for building data structures.

Lindahl and Sagonas \cite{LindahlSagonas2006,SagonasSilvaTamarit2013} designed
a success typing system for Erlang which infers types with a constraint-based
algorithm. Types are drawn from 
a finite lattice that encompasses types for various atoms (symbols,
numbers, strings, etc), functions,
tuple and list constructions, unions, and a type \textit{any} that
subsumes all other types. One of the major
goals of their approach is the ability to automatically generate
documentation for functions from the inferred success types. This goal
requires small, readable types, which are guaranteed by the finiteness
of the lattice. Types for data structures are made finite by cutting
off at a certain depth bound. A concrete
example shows where this boundedness leads to approximation.

Many Erlang programming idioms rely on named tuples, that is, tuples where
the first component is an atom and the remaining components contain
associated data as in
\lstinline[language=erlang,breaklines=true]|{book,"Hamlet","Shakespeare"}|.
One can view named tuples as named constructors:
\lstinline[language=erlang,breaklines=true]|book("Hamlet","Shakespeare")|.
Named tuples can be nested arbitrarily and created dynamically.

Lindahl and Sagonas' algorithm misses some definite errors based on nested
named tuples, as can be seen by the following example. Here is an
implementation of a list length function returning the zero constructor and
succ constructor instead of the built-in integers.\footnote{The
left-hand side pattern \lstinline|[]| matches the empty list and the
pattern \lstinline,[_|XS], matches a list with arbitrary head and tail
bound to \lstinline{XS}.}
\begin{lstlisting}[language=erlang,firstline=5,lastline=6]
length([]) -> {zero};
length([_|XS]) -> {succ, length(XS)}.
\end{lstlisting} The
Dialyzer\footnote{The DIscrepancy AnalYZer for ERlang programs, an
implementation of Lindahl and Sagonas' algorithm.
\url{http://www.erlang.org/doc/man/dialyzer.html}} infers the following
success type for length:
\[\mathrm{length} : [any] \to \mathrm{zero} \cup \mathrm{succ}(\mathrm{zero} \cup
\mathrm{succ}(\mathrm{zero}) \cup \mathrm{succ}(any)) \]
The argument part of the success type, $[any]$, describes that applying length
to a non-list argument yields an error and applying it to a list of arbitrary
content might succeed or fail. The result part describes the return value as
either $\mathrm{zero}$ or as a nested tuple consisting of $\mathrm{succ}$ and
any value. The argument part is exact: There is no argument of type
$[any]$ for which length fails. However, the analyzer restricts tuples to a
nesting depth of three levels.

To illustrate the problem with this approximation, consider the
function \texttt{check} that pattern 
matches on a nest of named tuples, which cannot be created by the
\texttt{length} function. Applying the \texttt{check}
function to the result of \texttt{length} yields a definite error. However, the
standard setting of the Dialyzer does not detect this error.
\begin{lstlisting}[language=erlang,firstline=8,lastline=9]
check({succ,{succ,{succ,{foo}}}}) -> 0.
test() -> check(length([0,0,0,0])).
\end{lstlisting}

\subsection{Our Approach}

We focus on errors that include the creation and
destruction of data structures and thus consider programs that manipulate
constructor trees, only. Our approach describes input type and output type of
a function with different models. A success typing of a function comprises
a recursive type describing the possible outputs and of a crash condition as a
logical formula whose models are the crashing inputs of the function. This
approach yields a modular definition of success typings.

\paragraph{Contributions}

\begin{itemize}
    \item We propose a new formally defined view of success typing for a language
    with data structures. We represent the input and output
    types of a function differently and thus obtain a modular approach.
    \item Our approach is correct. We show preservation
    of types and crash condition during evaluation as well as failure
    consistency (i.e., 
    if our analysis predicts a crash, the evaluation crashes definitely).
    \item We give a prototype implementation of our approach.
\end{itemize}

\paragraph{Outline}

In \cref{sec:Language} we define syntax and semantics of a
constructor-based language. We introduce types and crash conditions
for expressions of this language in \cref{sec:TypeAndCrashCondition} followed by
an analysis that assigns types and crash conditions to expressions.
Afterwards, we show show the correctness of the analysis. We discuss practical
issues of our approach in \cref{sec:PracticalConsiderations}.
In \cref{sec:RelatedWork} we discuss related work and conclude in
\cref{sec:Conclusion}.

An extended version of this article, including proofs, is available
online~\cite{JakobThiemann2015}.

\section{Language}
\label{sec:Language}

We illustrate our approach using a higher-order call-by-value language
\Clangname that comprises of explicit recursion, integer values, $n$-ary
constructors, and pattern-matching to distinguish and destruct the previously
defined constructors. We draw these constructors from a fixed, finite, and
distinct ranked alphabet, that is, every constructor has a specific arity. We
will denote constructors by upper case letters $A,B,C,\dots$, and implicitly
specify their arity when creating constructor terms.

\paragraph{Syntax} Syntactically, the language \Clangname
(\cref{fig:SyntaxOfLang}) consists of values and expressions. A value \Cval is
either an integer literal $n$, a constructor term $\Cctorvn$ where $\Cctor$
has arity $n$ and $\Cval_i$ are values, a recursive unary\footnote{Multiple
arguments can be passed by wrapping them in a constructor.} function
$\Frec{f}{x}{\Cexp}$, or an explicit error \Cerr. An expression \Cexp is
either a value \Cval, an identifier $x$, a constructor term $\Cctoren$ where
$\Cctor$ has arity $n$ and $\Cexp_i$ are expressions, a function application
$\Fapp{\Cexp}{\Cexp}$, or a pattern-matching expression $\Fmatch{\Cexp}{P}$.
Within the possibly empty list of patterns $P$, a pattern $\Cctorxn\rightarrow
\Cexp$ consists of a constructor \Cctor with arity $n$, a list of variables
$x_1,\dots,x_n$, and a body expression $\Cexp$. For the list of empty patterns
we write \Cemptypattern and to append lists we write $[\Cctorxn\rightarrow
\Cexp] \append \Cpatterns$. We assume that the constructors in a list
of patterns occur at most once. We introduce an auxiliary definition \CvalNoFun
that represents values not containing functions. For constructor expressions
with arity zero, we omit parentheses.

\paragraph{Semantics} In \cref{fig:Semantics} we define the semantics of
\Clangname as a small-step operational semantics. We use $\cE$ to
describe expressions with holes \Chole and \textsc{Eval-Final} to evaluate
expressions containing only values as subexpressions. \textsc{Eval-Hole}
evaluates expressions by choosing holes. \textsc{SApp} defines recursive
function application by capture-avoiding substitution of the argument and
function symbol. The rule \textsc{SMatch} evaluates a constructor value
and a list of patterns if the the constructor value matches the first pattern.
If so, it extracts the values of the argument and substitutes the variables
for the corresponding values in the pattern's body expression. If the first
pattern in the list of patterns does not match the constructor value,
the rule \textsc{SMatchNext} applies and discards the first non-matching
pattern.

We explicitly define error creation and propagation as we want to detect
definite errors in our programs. Errors occur, if the expression at the first
argument of a function application is reduced to a non-function value or if a
pattern matching expression occurs with an empty list either because no
pattern matched or the list of patterns was initially empty. The former case,
non-function values in applications, is handled by the rule \textsc{SAppErr1}
and the latter case by the rule \textsc{SMatchErr} both reducing to the error
value \Cerr. Error propagation is handled by the rules \textsc{SMatchErr}, if
the argument to a pattern matching is an error, the rule \textsc{SAppErr2} if
the argument to a function application is an error, and the rule
\textsc{SMatchNextErr}, if a constructor contains an error as a subexpression.

\begin{figure}
	\begin{alignat*}{5}
	    &\Cval &&::= n \sep \Frec{f}{x}{\Cexp} \sep \Cctorvn \sep \Cerr\\
	    &\Cexp &&::= \Cval \sep x \sep \Cctoren \sep \Fapp{\Cexp}{\Cexp}
	    	\sep \Fmatch{\Cexp}{[\overline{\Cctor_i(\bar{x}) \rightarrow \Cexp_i}]}\\
	    &\CvalNoFun &&::= n \sep \Cctorvn
  	\end{alignat*}
	\caption{Syntax of \Clangname with values $\Cval$, expressions $\Cexp$ and non-function values $\CvalNoFun$.}
	\label{fig:SyntaxOfLang}
\end{figure}

\begin{figure}
	\begin{align*}
		\cE ::=\ & \ C(\Cval_1,\dots,\Cval_n,\Chole,\Cexp_1,\dots,\Cexp_m) \sep \Chole\ \Cexp \sep v\ \Chole
				\sep \Fmatch{\Chole}{[C_i(\bar{x})\rightarrow \Cexp_i]}
	\end{align*}
	\[\inferrule[Eval-Final]{\Fevals{\Cexp}{\Cexp'}}{\Cexp\hookrightarrow \Cexp'}\quad
	\inferrule[Eval-Hole]{\Cexp\hookrightarrow \Cexp'}{\cE[\Cexp]\hookrightarrow \cE[\Cexp']}\]

	\begin{alignat*}{3}
		&\Fsemanticrule{SApp}{&&\Fapp{(\Frec{f}{x}{\Cexp})}{\Cval}&&}{\Cexp [x\mapsto \Cval,
				f\mapsto\Frec{f}{x}{\Cexp}]}\\
		&\Fsemanticrule{SAppErr1}{&&\Fapp{\CvalNoFun}{\Cval}&&}{\Cerr}\\
		&\Fsemanticrule{SAppErr2}{&&\Fapp{(\Frec{f}{x}{\Cexp})}{\Cerr}&&}{\Cerr}\\
		&\Fsemanticrule{SMatch}{&&\Fmatch{\Cctorvnabbr}{[\Cctorxnabbr\to\Cexp,\dots]}&&}
						{\Cexp[\overline{x_i\mapsto \Cval_i}]}\\
		&\Fsemanticrule{SMatchErr}{&&\Fmatch{\Cerr}{[\dots]}&&}{\Cerr}\\
		&\Fsemanticrule{SMatchNext}{&&\Fmatch{\Cctorvnabbr}{[\Cctordxmabbr \to \Cexp] \append P}&&}
						{\Fmatch{\Cctorvnabbr}{P}}\\
		&\Fsemanticrule{SMatchNextErr}{&&\Fmatch{\Cctorvnabbr}{[\ ]}&&}{\Cerr}\\
		&\Fsemanticrule{SCtorErr}{&&\Cctor(\Cval_1,\dots,\Cval_n,\Cerr,\Cexp_1,\dots,\Cexp_m)&&}{\Cerr}
	\end{alignat*}
\caption{Small-step operational semantics for \Clangname.}
\label{fig:Semantics}
\end{figure}

\section{Type and Crash Condition}
\label{sec:TypeAndCrashCondition}

The basic notion of our formalization is a type \Ctype that
represents trees created from constructors \Cctor on a type
level. Furthermore, we represent function values using recursive
types. To formalize success types, we represent the possible outputs
of a function and the valid inputs of a function differently, thus
resulting in a non-standard function type definition where the possible
outputs are represented using a type \Ctype and the possible inputs
are represented using a crash condition \Ccc. Types and crash conditions
are defined mutually in \cref{fig:DefinitionsOfTypesAndCrashConditions}.
Intuitively, a crash condition for a function is a logical formula whose
models are types. These types describe inputs that definitely crash the
function.

Types \Ctype comprise of type variables \Ctvar, an equi-recursive function type
written \Ftfun{\Ctvar}{\Ctype}{\Ccc} that includes a type variable \Ctvar representing
the function's argument, a return type \Ctype, and a crash condition \Ccc
indicating when the function definitely crashes. Furthermore, we define
a constructor type \Ctype that captures the types of a constructor expression,
a union type $\Ctype \cup \Ctype$, an integer type \Ctint, and the empty type
\Ctbot that has no values. We define two operators that work on types: a
type-level function application $\Ftapp{\Ctype}{\Ctype}$, and a projection
function for constructor types $\Ftget{\Ctype}{\Cctor}{i}$ that projects the
$i$th component of a type \Ctype if it is a constructor type \Cctor. The
semantics of these operators is defined in \cref{fig:TypeAndCrashConditionOperators}.
In our definition, the fix-point formulation $\mu X$ only occurs together
with a function type definition. The type operators are always implicitly
applied.

Crash conditions \Ccc are defined as atoms true $\Ctt$ and false $\Cff$,
intersection $\Ccc\vee\Ccc$ and conjunction $\Ccc\wedge\Ccc$, predicates over
types $\Fhasctor{\Cctor}{\Ctype}$ symbolizing that a type \Ctype can be a
constructor \Cctor, $\Fhasnctor{\Cctor}{\Ctype}$ symbolizing that a type
\Ctype is not a constructor type \Cctor, and $\Fhasnfun{\Ctype}$ symbolizing
that \Ctype is not a function. Furthermore, in
\cref{fig:TypeAndCrashConditionOperators} we define an operator
$\Fcc{\Ctype}{\Ctype}$ that describes a crash-condition-level function
application. Again, the crash condition operator is implicitly applied.

An interpretation $\Cinter$ is a mapping of type variables to types. An
interpretation of a type $\Ftinter{\tau}$ is a set of types as specified in
\cref{fig:DefinitionOfAnInterpretation}.

\begin{figure}
\begin{alignat*}{3}
    &\Ftinter{\alpha} &&= \{\Cinter(\alpha)\}\\
    &\Ftinter{\Ftfun{\Ctvar}{\Ctype}{\Ccc}} &&= \{\Ftfun{\Ctvar}{\Ctype'}{\Ccc}
        \mid \Ctype'\in\Ftinterp{\Ctype}, \Cinterp = \Cinter \setminus \{\alpha\} \}\\
    &\Ftinter{\Ctctortn} &&= \{\Cctor(\Ctype_1',\cdots,\Ctype_n') \mid \Ctype_i'\in\Ftinter{\Ctype_i})\\
    &\Ftinter{\Ctype_1 \cup \Ctype_2} &&= \Ftinter{\Ctype_1} \cup \Ftinter{\Ctype_2}\\
    &\Ftinter{\Ctint} &&= \{\Ctint\}\\
    &\Ftinter{\Ctbot} &&= \{\}\\
\end{alignat*}
    \caption{Definition of an interpretation \Cinter on a type \Ctype.}
    \label{fig:DefinitionOfAnInterpretation}
\end{figure}

In \cref{fig:DefinitionOfTheEntailmentRelation} we
recursively define an entailment relation $\Cinter \vDash \Ccc$ for an interpretation $\Cinter$
and a crash condition $\Ccc$.

\begin{figure}
\begin{alignat*}{3}
    \Cinter &\vDash \Ctt\\
    \Cinter &\nvDash \Cff\\
    \Cinter &\vDash \Ccc_1 \vee \Ccc_2 &&\iff \Cinter \vDash \Ccc_1 \vee \Cinter \vDash \Ccc_2\\
    \Cinter &\vDash \Ccc_1 \wedge \Ccc_2 &&\iff \Cinter \vDash \Ccc_1 \wedge \Cinter \vDash \Ccc_2\\
    \Cinter &\vDash \Fhasctor{\Cctor}{\Ctype} &&\iff \exists \Cctor(\overline{\Ctype}) \in \Ftinter{\Ctype}\\
    \Cinter &\vDash \Fhasnctor{\Cctor}{\Ctype} &&\iff \nexists (\Cctor(\overline{\Ctype})) \in \Ftinter{\Ctype}\\
    \Cinter &\vDash \Fhasnfun{\Ctype} &&\iff \nexists (\Ftfun{\Ctvar}{\Ctype}{\Ccc}) \in \Ftinter{\Ctype}
\end{alignat*}
    \caption{Definition of the entailment relation $\Cinter \vDash \Ccc$.}
    \label{fig:DefinitionOfTheEntailmentRelation}
\end{figure}

\begin{example}
    \label{ex:length}
    We take the length function of lists as an example using constructors
    $\Cctor_{nil}, \Cctor_{zero}, \Cctor_{succ}$, and $\Cctor_{cons}$ with
    arities zero, zero, one, and two, respectively.
    \begin{align*}
        \Frec{len}{x}{\Fmatch{x}{[\Cctor_{nil}\rightarrow \Cctor_{zero}, \Cctor_{cons}(x_1,x_2)\rightarrow \Cctor_{succ}(\Fapp{len}{x_2})]}}
    \end{align*}
	A possible function type for the length function is
	\begin{multline*}
    \Ctype_{len}=\Ftfun{\Ctvar}{\\ \Cctor_{zero}\cup \Cctor_{succ}(\Ftapp{X}{\Ftget{\Ctvar}{\Cctor_{cons}}{2}})}
        {\Fhasnctor{\Cctor_{nil}}{\Ctvar} \wedge
              \left(\left(\Fhasctor{\Cctor_{cons}}{\Ctvar} \wedge \Fcc{X}{\Ftget{\Ctvar}{\Cctor_{cons}}{2}} \right)
              \vee \Fhasnctor{\Cctor_{cons}}{\Ctvar}\right)
              }
    \end{multline*}
	whose type is recursively entwined with its crash condition. The derivation
    of this type is described in \cref{ssec:Analysis}.
	We extract the crash condition that still makes use of $\Ctype_{len}$ via $X$
    and get a logical formula with free variable $\Ctvar$
	\[\Ccc_{len}=\Fhasnctor{\Cctor_{nil}}{\Ctvar} \wedge
              \left(\left(\Fhasctor{\Cctor_{cons}}{\Ctvar} \wedge \Fcc{\Ctype_f}{\Ftget{\Ctvar}{\Cctor_{cons}}{2}} \right)
              \vee \Fhasnctor{\Cctor_{cons}}{\Ctvar}\right)
		  \]
	that symbolizes when the function crashes.
	For example the following interpretation (amongst many others)
	\[\Cinter=\{\alpha \mapsto \bigcup \{
        \Fcc{
            \left(\Ftfun{\Ctvar}{
                \Cctor_{zero} \cup \Cctor_{cons}\left(\Ctype,\Ftapp{X}{\Ctvar}\right)
                }
                {\Cff}
                \right)
            }{\Cctor_{unused}} \mid \Ctype \in \Ctypes\}\]
	entails the crash condition: $\Cinter \vDash \Ccc_{len}$. Here, $\Cctor_{unused}$ is
    only needed as a dummy argument to the type-level function.
    When implicitly applying the type operators, we end up with
    the infinite type\footnote{For the sake of a simpler type syntax, this type cannot
    be represented using our type syntax directly. We always have to use type-level applications.}
    \[\{\mu X.\Cctor_{zero} \cup \Cctor_{cons}(\Ctype,X) \mid \Ctype \in \Ctypes\}\]
    This type represents all lists not ending with
	a nil but with a zero.
\end{example}

\begin{figure}
	\begin{alignat*}{5}
		&\Ctype  &&::= \Ctvar \sep \Ftfun{\Ctvar}{\Ctype}{\Ccc}
				\sep \Ctctortn \sep \Ctype \cup \Ctype \sep \Ctint \sep \Ctbot
				\sep \Ftapp{\Ctype}{\Ctype}
				\sep \Ftget{\Ctype}{\Cctor}{i}\\
		&\Ccc &&::= \Cff \sep \Ctt \sep \Ccc\vee\Ccc \sep \Ccc\wedge\Ccc
			\sep \Fhasctor{\Cctor}{\Ctype} \sep \Fhasnctor{\Cctor}{\Ctype} \sep \Fhasnfun{\Ctype} \sep
			\Fcc{\Ctype}{\Ctype}
	\end{alignat*}
	\caption{Definition of types $\Ctype$ and crash conditions $\Ccc$.}
	\label{fig:DefinitionsOfTypesAndCrashConditions}
\end{figure}

\begin{figure}
		\begin{mathpar}
			\Ftapp{\Ctype_1}{\Ctype_2} =
				\begin{cases}
					\Ctype_b[\Ctvar\mapsto \Ctype_2, X\mapsto \Ctype_1]
							&\textnormal{if } \Ctype_1=\Ftfun{\Ctvar}{\Ctype_b}{\Ccc}\\
					\Ftapp{\Ctype_1}{\Ctype_2}
							&\textnormal{if } \Ctype_1=\Ctvar\\
					\Ftapp{\Ctype_{11}}{\Ctype_2} \cup \Ftapp{\Ctype_{12}}{\Ctype_2}
							&\textnormal{if } \Ctype_1=\Ctype_{11}\cup \Ctype_{12}\\
					\Ftapp{\Ctype_1}{\Ctype_{21}} \cup \Ftapp{\Ctype_1}{\Ctype_{22}}
							&\textnormal{if } \Ctype_2=\Ctype_{21}\cup \Ctype_{22}\\
					\Ctbot
							&\textnormal{otherwise}
				\end{cases}

			\and

			\Ftget{\Ctype}{\Cctor}{i} =
				\begin{cases}
					\Ctype_i
							& \textnormal{if}\ \Ctype = \Ctctortn, 1\leq i\leq n\\
					\Ftget{\Ctype}{\Cctor}{i} & \textnormal{if}\ \Ctype_0 = \Ctvar\\
					\Ctbot & \textnormal{otherwise}
				\end{cases}

			\and

			\Fcc{\Ctype_1}{\Ctype_2} =
				\begin{cases}
					\Ccc[\Ctvar\mapsto \Ctype_2, X\mapsto \Ctype_1]
							&\textnormal{if } \Ctype_1=\Ftfun{\Ctvar}{\Ctype_b}{\Ccc}\\
					\Fcc{\Ctype_1}{\Ctype_2}
							&\textnormal{if } \Ctype_1=\Ctvar\\
					\Fcc{\Ctype_{11}}{\Ctype_2} \cup \Fcc{\Ctype_{12}}{\Ctype_2}
							&\textnormal{if } \Ctype_1=\Ctype_{11}\cup \Ctype_{12}\\
					\Fcc{\Ctype_1}{\Ctype_{21}} \cup \Fcc{\Ctype_1}{\Ctype_{22}}
							&\textnormal{if } \Ctype_2=\Ctype_{21}\cup \Ctype_{22}\\
					\Ctt
							&\textnormal{otherwise}
				\end{cases}
		\end{mathpar}
	\caption{Type and crash condition operators.}
	\label{fig:TypeAndCrashConditionOperators}
\end{figure}

Before introducing the analysis that assigns types and crash conditions to
expressions, please note that the question of entailment is not decidable in
general.
\begin{lemma}
    It is undecidable whether for an arbitrary crash condition $\Ccc$ there exists
    an interpretation \Cinter such that $\Cinter \vDash \Ccc$.
    \label{lem:UndecidableSatisfiability}
\end{lemma}
We discuss possible solutions to this problem in \cref{sec:PracticalConsiderations}.

\subsection{Analysis}
\label{ssec:Analysis}

We present our analysis as a type system using a judgment
$\trule{\Cexp}{\Ctype}{\Ccc}$ that relates a type variable environment \Cenv,
an expression \Cexp, a type \Ctype of the expression, and a crash condition
\Ccc characterizing when the expression crashes. We define the derivation
rules in \cref{fig:DerivationRulesForTheTypesAndCrashConditions}.

The rule \textsc{T-Rec} derives a recursive function type for a recursive
function expression by inferring the body's type and crash condition using
type variables for the argument and a recursive type formulation for recursive
calls. For a function application (\textsc{T-FunApp}) we infer types and crash
conditions for both the callee $\Cexp_1$ and the argument $\Cexp_2$ . The
result type of the function application is the type-level application of the
types of the callee and the argument. The function application can crash if
either $\Cexp_1$ or $\Cexp_2$ crashes, $\Cexp_1$ is not a function, or the
application itself crashes. The latter is symbolized by a crash condition-level
function application. The rule \textsc{T-Identifier} derives the type of
a variable from the environment and never crashes. An error value $\Cerr$ has
type $\Ctbot$ and always crashes (\textsc{T-Error}). In rule
\textsc{T-Constructor}, a constructor expression has a constructor type with
the types of its arguments inferred recursively. A constructor crashes if one
of its arguments crashes. Integer literals are handled by \textsc{T-Integer} and
always have type \Ctint and never crash.

For the pattern matching expression, the type is described by the union of the
types of the expression in the patterns. The crash condition is described by
the crash condition of the expression to match and the crash conditions of the
cases. The crash conditions of the cases are built using an auxiliary
judgment: $\pruleenv{\Ccc_m;\Ctype_0}{\Cpatterns}{\Ctype_p}{\Ccc_p}$ where
$\Ccc_m$ describes the crash conditions accumulated so far, $\Ctype_0$
describes the type of the expression to match, $\Cpatterns$ the list of
patterns which are traversed and $\Ctype_p$ the union of the types of the
pattern case's body expression. The type and crash condition of a pattern list
is created by two rules: if the pattern list is empty, we return the bottom
type and the crash condition accumulated to far.  If the pattern list is
non-empty, we create the type of the current body expression by binding the
variables defined in the pattern and inductively applying the derivation.  The
current expression can crash, if either the pattern matches
($\Fhasctor{C}{\Ctype_0}$) and the body expression crashes, or if the pattern
does not match at all.

Additionally, we define a subtyping relation $\leq: \Ctype \times \Ctype$ in
\cref{fig:SubtypingRules} The relation is standard, except for the rule
\textsc{S-Fun}, which requires a logical implication of the crash conditions.

The (output) types derived for an expression are over-approximations whereas
the crash conditions describe the possible crashes exactly. The interplay
of types and crash conditions ends up with definite errors, because the predicate
$\Fhasnctor{\Cctor}{\Ctype}$ describes the question whether it is not possible
that the type \Ctype is a constructor \Cctor, and similarly for the predicate
$\Fhasnfun{\Ctype}$.

\begin{figure}
	\begin{mathpar}
		\trulerec\and
		\truleapp\and
		\truleid\and
		\truleerr\and
		\trulector\and
		\trulepm\and
		\truleinteger\and
		\trulepmnext\and
		\trulepmempty
	\end{mathpar}

	\caption{Derivation rules for the types and crash conditions.}
	\label{fig:DerivationRulesForTheTypesAndCrashConditions}
\end{figure}

\begin{figure}
	\begin{mathpar}
		\trulesubbot\and
		\trulesubunion\and
		\trulesubrefl\and
		\trulesubctor\and
		\trulesubfun\and
		\trulesubgeneral
	\end{mathpar}
	\caption{Subtyping rules.}
	\label{fig:SubtypingRules}
\end{figure}

\subsection{Properties}
\label{sec:Properties}

To justify our analysis, we prove the preservation of types and crash
conditions and the correctness. To do so, we need several auxiliary lemma.

Weakening allows the introduction of a fresh type variable into the
type environment without changing anything.
\begin{lemma}[Weakening]For expressions $\Cexp$, types $\Ctype$, $\Ctype_y$, and $\Ctype_0$, an
	identifier $y$, conditions $\Ccc$ and $\Ccc_0$, and an environment $\Cenv$,
	the following holds:
	\begin{enumerate}
	  \item If $\trule{\Cexp}{\Ctype}{\Ccc}$ and $y\notin\Fdom{\Cenv}$ then
	  	$\truleenv{y:\Ctype_y}{\Cexp}{\Ctype}{\Ccc}$
	  \item If $\pruleenv{\Ccc_0;\Ctype_0}{\Cpatterns}{\Ctype}{\Ccc}$
	  and $y\notin\Fdom{\Cenv}$ then
	    $\pruleenvown{\Ccc_0;\Ctype_0}{\Cenv,y:\Ctype_y}{\Cpatterns}{\Ctype}{\Ccc}$.
	\end{enumerate}
	\label{lem:Weakening}
\end{lemma}

\newcommand{\amato}[1]{#1[\Ctvar\mapsto \Ctype_\Ctvar]}

The next lemma shows that we can replace a type variable \Ctvar within an
environment by a concrete type \Ctvar if we replace all occurrences of the
type variable in the resulting type and crash condition. We need this lemma
when working with the type-level function application.
\begin{lemma}[Consistency of type substitution]
For a well-formed environment $\Cenv$, an identifier $y$, a type variable $\Ctvar$,
an arbitrary expression $\Cexp$, types $\Ctype$ and $\Ctype_\Ctvar$, conditions $\Ccc$ and $\Ccc_0$ the
following holds:
	\begin{enumerate}
		\item If $\truleenv{y:\Ctvar}{\Cexp}{\Ctype}{\Ccc}$ then
		$\truleenvown{\amato{(\Cenv,y:\Ctvar)}}{\Cexp}{\amato{\Ctype}}{\amato{\Ccc}}$
		\item If $\pruleenvown{\Ccc_0;\Ctype_0}{\Cenv,y:\Ctvar}{\Cpatterns}{\Ctype}{\Ccc}$ then
		$\pruleenvown{\amato{\Ccc_0};\amato{\Ctype_0}}{\amato{(\Cenv,y:\Ctvar)}}{\Cpatterns}{\amato{\Ctype}}{\amato{\Ccc}}$.
	\end{enumerate}
	\label{lem:ConsistencyOfTypeSubstitution}
\end{lemma}

\newcommand{\xmato}[1]{#1[{y\mapsto \Cval}]}

\Cref{lem:ConsistencyOfValueSubstitution} shows that we can substitute
a variable $y$ in an expression \Cexp with a value of the same type without
changing the type and crash condition of the whole expression. This lemma
is needed for the type-level function applications later.
\begin{lemma}[Consistency of value substitution]
	For an environment $\Cenv$, an identifier $y$, types $\Ctype$ and $\Ctype_y$, an
	expression $\Cexp$, conditions $\Ccc$ and $\Ccc_0$, and a value $\Cval$, the following holds
	\begin{enumerate}
		\item If $\truleenv{y: \Ctype_y}{\Cexp}{\Ctype}{\Ccc}$ and $\trule{\Cval}{\Ctype_y}{\Cff}$ then
	          $\trule{\xmato{\Cexp}}{\Ctype}{\Ccc}$.
		\item If $\pruleenvown{\Ccc_0;\Ctype_0}{\Cenv,y:\Ctype_y}{\Cpatterns}{\Ctype}{\Ccc}$ and $\trule{\Cval}{\Ctype_y}{\Cff}$ then $\pruleenvown{\Ccc_0;\Ctype_0}{\Cenv}{\xmato{\Cpatterns}}{\Ctype}{\Ccc}$.
	\end{enumerate}
	\label{lem:ConsistencyOfValueSubstitution}
\end{lemma}

The next lemma shows that the analysis is designed such that after a successful pattern matching,
the crash conditions of the remaining pattern's body expressions cannot be satisfied anymore. The reason
is that $\Ccc_0$ in the rules \textsc{T-Pattern-Next} and \textsc{T-Pattern-Empty} influences
the resulting crash condition of the whole expression.
\begin{lemma}[Unsatisfiability after matching patterns]
	For an environment $\Cenv$, types $\Ctype$ and $\Ctype'$, and conditions $\Ccc'$
	it holds that if $\pruleenv{\Cff,\Ctype_0}{\Cpatterns}{\Ctype'}{\Ccc'}$ then $\nvDash \Ccc'$.
	\label{lem:UnsatisfiabilityAfterMatchingPatterns}
\end{lemma}

Finally, we can establish the preservation theorem for our type system.
\begin{theorem}[Preservation of types and crash conditions]
	If $\trule{\Cexp}{\Ctype}{\Ccc}$, $\Cexp \hookrightarrow \Cexp'$ and $\trule{\Cexp'}{\Ctype'}{\Ccc'}$
	then $\subtype{\Ctype}{\Ctype'}$ and $\Ccc \leftrightarrow \Ccc'$.
	\label{thm:Preservation}
\end{theorem}

Furthermore, we show that our analysis is sound: if the crash conditions report
an error, then there is either an error or the evaluation does not terminate.

\begin{theorem}[Failure]
	If $\forall\cT,\cV,\Cenv$ and $\forall x\in\Fdom{\Cenv}$: $\vdash \cV(x) : \cT(\Cenv(x))$
    $\vDash\cT(\Ccc)$, and
    \begin{enumerate}
        \item $\trule{\Cexp}{\Ctype}{\Ccc}$, then $\cV(\Cexp)\hookrightarrow^* \Cerr$ or $\cV(\Cexp)\!\Uparrow$.
        \item $\pruleenv{\Ccc_0;\Ctype_0}{[\Cctor_1(\bar{x}\rightarrow \Cexp_1),\dots,\Cctor_n(\bar{x}\rightarrow \Cexp_n)]}{\Ctype}{\Ccc}$ and
        a value \Cctorvn with $\truleenvown{\Cenv'}{\Cctorvn}{\Ctctortn}{\Cff}$, then
        either
        \begin{itemize}
            \item $\forall i: \Cctor_i \neq \Cctor$
            \item or $\exists i: \Cctor_i = \Cctor$
                and ($\cV'(\Cexp_i) \hookrightarrow^* \Cerr$ or $\cV(\Cexp_1)\!\Uparrow$).
        \end{itemize}
    \end{enumerate}
	\label{thm:Failure}
\end{theorem}

\section{Practical Considerations}
\label{sec:PracticalConsiderations}

We have shown that success typing is an instance of falsification and thus
allows the detection of definite errors. However, as shown by
\cref{lem:UndecidableSatisfiability}, the satisfiability of $\Ccc$ is
undecidable in general. Thus a direct algorithmic solution cannot exist. We
implemented\footnote{\OnlineLocation} a version of the analysis that imposes a
user-definable limit of $k$ iterations on the unfolding operations described
in the operators in \cref{fig:TypeAndCrashConditionOperators} and can thus
check for errors up to depth $k$.

\begin{example}
    An example for a yet problematic combination of type and crash condition we cannot solve
    at the moment is the following: We create a function that generates an infinite
    list and apply the resulting stream on the list length function.

    With the list generator's type
    \begin{align*}
        \Ctype_{gen} =\ & \Ftapp{(\Ftfun{\alpha}{\Cctor_{cons}(\Cctor_{zero},\Ftapp{X}{\alpha})}{\Cff})}{\Cctor_{unused}}\\
        =\ & \Cctor_{cons} ( \Cctor_{zero},\Ftapp{\Ctype_{gen}}{\Cctor_{unused}})
    \end{align*}
    and the list's type from \cref{ex:length} the application of the stream to the length function
    has the following crash condition after type and crash condition operators are applied once (before
    substitution):
    \begin{multline*}
       \left(\Fhasnctor{\Cctor_{nil}}{\Ctvar} \wedge
              \left(\left(\Fhasctor{\Cctor_{cons}}{\Ctvar} \wedge \Fcc{X}{\Ftget{\Ctvar}{\Cctor_{cons}}{2}} \right)
              \vee \Fhasnctor{\Cctor_{cons}}{\Ctvar}\right)\right)\\
       [
        \alpha \mapsto \Cctor_{cons}(\Cctor_{zero},\Ftapp{\Ctype_{gen}}{\Cctor_{unused}})
       ]
    \end{multline*}
	After performing the substitution, we can evaluate the predicates that only look finitely deep into
    their argument. When we apply type and crash condition operators again,
	we end up on the same crash condition.
    Although we reach a fix point in this case, this is of course not the case in general.
\end{example}

To solve this problem in general, we need to find an approximation for the crash condition formula.
As we only want to find definite errors, our approximation has to be an under-approximation.
However, finding a good under-approximation, is yet an open problem.

When we view the output type of a functions as a constructor tree, we can represent
it as a higher-order tree grammar, as is proposed by Ong and Ramsey~\cite{OngRamsey2011}.
The (approximated) crash condition of a function can be represented as a tree automaton. As the model
checking of tree automata and higher-order tree grammars is decidable~\cite{Ong2006}
we have some means of finding definite errors.


\section{Related Work}
\label{sec:RelatedWork}

The idea of finding definite errors in programs is quite old and several
approaches exist.

Constraint-based analyses to detect must-information can be found in Reynolds
\cite{Reynolds1968} where he describes a construction of recursive set
definitions for LISP programs that are ``a good fit to the results of a
function''. However, the goal of the paper was to infer data structure
declarations and not to find errors. The constraint-based analysis of Lindahl
and Sagonas' \cite{LindahlSagonas2006} is a modular approach similar to ours,
but does not account for data structures of arbitrary depth but instead uses
k-depth abstraction as we do in our current implementation. Furthermore, the
approach of Lindahl and Sagonas uses union types that are widened after a
fixed size limit. These limits are to establish small and readable types
whereas we focus on exact tracking of values.

Soft typing, presented by Cartwright and Fagan~\cite{CartwrightFagan1991long}
detects suspicious expressions in a program, i.e., expressions that cannot be
verified to be error-free, and adds run-time checks. Although the idea of not
rejecting working programs is the same, our approach requires no changes in
existing programs as we only assume programs to contain errors if we can proof
it.

The line of work of Vaziri et al.~\cite{JacksonVaziri2000,DolbyVaziriTip07}
focuses on imperative first-order languages and uses user-defined specifications
given in the Alloy language to state the
intention of a function and then checks the implementation against its
specification. Although they explicitly mention unbounded data structures in
their approach, only instances up to a number of heap cells and loop
iterations are considered. In contrast to our approach, they require
user-defined annotations. A similar framework~\cite{Taghdiri2004} removes the chore
to define annotations and only requires the user to provide a property to be
checked. Their abstraction refines specifications that describe the behavior
of procedures and thus creates a refinement-based approach that ensures that
no spurious errors appear if the analysis halts.

Different approaches for definite error detections are presented by Ball et
al.~\cite{BallKupfermannYorsh2005} and Kroening and Weissenbacher
\cite{KroeningWeissenbacher2010} for imperative first-order languages in a
Hoare-style way. However, a comparison to our approach is difficult because
they rely on a transition system to model the behavior of programs whereas we
use a type system.

\section{Conclusion}
\label{sec:Conclusion}

We presented a new formal approach to success typings for a constructor-based
higher-order language using different representations for the input and output
type of a function. We proved that our formulation of success typings is a
falsification in the sense that it only reports definite errors. We presented
a prototype implementation that checks for errors up to a user-defined bound.

In future we want to look at means to model check (type) trees
\cite{Ong2006,OngRamsey2011} with logical formula represented as higher-order
tree grammars and tree automata, respectively. Thus, we hope to (partly) remove
the $n$-bound of current approaches.

\paragraph{Acknowledgments.} This work has been partially supported by the
German Research Foundation (Deutsche Forschungsgemeinschaft, DFG) within the
Research Training Group 1103 (Embedded Microsystems).


\input{literature}

\ifbool{IsExtVers}{

\clearpage

\appendix

\section{Proofs}
\label{sec:Proofs}

For the sake of completeness we enlist the substitution of variables
and of type variables in \cref{fig:SubstitutionOfVariables,fig:SubstitutionOfTypeVariables},
respectively.

\begin{figure}
	\begin{align*}
		x[x\mapsto \Cval] &= \Cval\\
		y[x\mapsto \Cval] &= y\\
		n[x\mapsto \Cval] &= n\\
		C(\bar{\Cexp})[x\mapsto \Cval] &= C(\bar{\Cexp}[x\mapsto \Cval])\\
		(\Fapp{\Cexp}{\Cexp})[x\mapsto \Cval] &= \Fapp{\Cexp[x\mapsto \Cval]}{\Cexp[x\mapsto \Cval]}\\
		([\Cctor_i(\bar{y})\rightarrow \Cexp])[x\mapsto \Cval] &= [\Cctor_i(\bar{y})\rightarrow \Cexp[x\mapsto \Cval]] &x\notin\{\bar{y}\}\\
		\Frec{f}{y}{\Cexp}[x\mapsto \Cval] &= \Frec{f}{y}{e[x\mapsto \Cval]} &x\notin \{f,y\}\\
		\Frec{f}{y}{\Cexp}[x\mapsto \Cval] &= \Frec{f}{y}{e} &x\in \{f,y\}
	\end{align*}
	\caption{Substitution of variables.}
	\label{fig:SubstitutionOfVariables}
\end{figure}

\begin{figure}
	\newcommand{\amat}{\Ctvar\mapsto \Ctype}
	\newcommand{\subamat}[1]{#1[\Ctvar\mapsto \Ctype]}
	\centering
	\begin{minipage}{.49\linewidth}
		\begin{align*}
			\Ctvar[\amat] &= \Ctype\\
			\beta[\amat] &= \beta\\
			\Ctbot[\amat] &= \Ctbot\\
			\Cctor(\bar{\Ctype})[\amat] &= \Cctor(\overline{\Ctype[\amat]})\\
			(\Ctype_1\cup \Ctype_2)[\amat] &= \Ctype_1[\amat]\cup \Ctype_2[\amat]\\
			\Ftfun{\Ctvar}{\Ctype}{\Ccc}[\amat] &= \Ftfun{\Ctvar}{\Ctype}{\Ccc}\\
			\Ftfun{\beta}{\Ctype}{\Ccc}[\amat] &= \Ftfun{\beta}{\Ctype[\amat]}{\Ccc[\amat]}\\
			\Ctint[\amat] &= \Ctint\\
			\Fapp{\Ctype_1}{\Ctype_2}[\amat] &= \Fapp{\Ctype_1[\amat]}{\Ctype_2[\amat]}\\
			\Ftget{\Ctype}{\Cctor}{i}[\amat] &= \Ftget{(\Ctype[\amat])}{\Cctor}{i}
		\end{align*}
	\end{minipage}
	\begin{minipage}{.49\linewidth}
		\begin{align*}
			\subamat{\Cff} &= \Cff\\
			\subamat{\Ctt} &= \Ctt\\
			\subamat{\Ccc_0 \wedge \Ccc_1} &= \subamat{\Ccc_0} \wedge \subamat{\Ccc_1}\\
			\subamat{\Ccc_0 \vee \Ccc_1} &= \subamat{\Ccc_0} \vee \subamat{\Ccc_1}\\
			\subamat{\Fhasctor{\Cctor}{\Ctype}} &= \Fhasctor{\Cctor}{(\subamat{\Ctype})}\\
			\subamat{\Fhasnctor{\Cctor}{\Ctype}} &= \Fhasnctor{\Cctor}{(\subamat{\Ctype})}\\
			\subamat{\Fhasnfun{\Ctype}} &= \Fhasnfun{(\subamat{\Ctype})}\\
			\subamat{\Fcc{\Ctype_1}{\Ctype_2}} &= \Fcc{\subamat{\Ctype_1}}{\subamat{\Ctype_2}}
		\end{align*}
	\end{minipage}\\
	\caption{Substitution of type variables.}
	\label{fig:SubstitutionOfTypeVariables}
\end{figure}

\begin{proof}[\cref{lem:UndecidableSatisfiability}]
  As types and crash conditions contain type-level functions, applications, and
  constructors, the crash conditions are Turing-complete and thus satisfiability
  is not decidable.
\end{proof}

\begin{proof}[\cref{lem:Weakening}]
Proof by induction on the derivation of the analysis. 
\begin{description}
	\item\case{T-Integer} Let $\Cenv'=\Cenv,y:\Ctype_y$. Then,
	$\truleenvown{\Cenv'}{n}{\Ctint}{\Cff}$ trivially holds by \textsc{T-Integer}.
	\item\case{T-Identifier} We assume $\trule{x}{\Ctype}{\Cff}$ and $y\notin\Fdom{\Cenv}$.
	Then, we have two subcases:
	\begin{description}
		\item\subcase{$x=y$} By inversion, we get $\Cenv(x)=\Cenv(y)=\Ctype$. Thus,
		$y\in\Fdom{\Cenv}$ which contradicts our assumption.
		\item\subcase{$x\neq y$} By inversion, we get $\Cenv(x)=\Ctype$. Let
		$\Cenv'=\Cenv,y:\Ctype_y$.  Then, $\Cenv'(x)=\Ctype$ still holds and we apply
		T-Identifier and conclude $\truleenvown{\Cenv'}{x}{\Ctype}{\Cff}$.
	\end{description}
	\item\case{T-Constructor} By assumption we have both
	\begin{align*}
        \trule{\Cctorenabbr}{\Ctctortnabbr}{\vee \bar{\Ccc}}\\
        y\notin\Fdom{\Cenv}
    \end{align*}
    Using inversion we get
    \[\trule{\Cexp_i}{\Ctype_i}{\Ccc_i}\]
	for $i=1,\dots,n$.  On each of these judgements, we can apply the induction
	hypothesis, and deduce \[\truleenv{y:\Ctype_y}{\Cexp_i}{\Ctype_i}{\Ccc_i}\]
    Finally, we apply \textsc{T-Constructor} and conclude that
	\[\truleenv{y:\Ctype_y}{\Cctorenabbr}{\Ctctortnabbr}{\bigvee \bar{\Ccc}}\]
    holds.
	\item\case{T-FunApp} We assume
	\begin{align*}
        &\trule{\Fapp{\Cexp_1}{\Cexp_2}}{\Ftapp{\Ctype_1}{\Ctype_2}}{\Fcc{\Ctype_1}{\Ctype_2} \vee \Ccc_1
	   \vee \Ccc_2 \vee \Fhasnfun{\Ctype_1}}\\
        &y\notin\Fdom{\Cenv}
    \end{align*}
    By inversion
	we get \[\trule{\Cexp_i}{\Ctype_i}{\Ccc_i}\] for $i=1,2$ and apply the induction
	hypothesis on both of them. With \[\truleenv{y:\Ctype_y}{\Cexp_i}{\Ctype_i}{\Ccc_i}\]
    we
	can apply \textsc{T-FunApp} and conclude
	\[\truleenv{y:\Ctype_y}{\Fapp{\Cexp_1}{\Cexp_2}}{\Ftapp{\Ctype_1}{\Ctype_2}}{\Fcc{\Ctype_1}{\Ctype_2} \vee
	\Ccc_1 \vee \Ccc_2 \vee \Fhasnfun{\Ctype_1}}\]
	\item\case{T-Rec} We assume
    \begin{align*}
        &\trule{\Frec{f_r}{x_r}{e}}{\Frec{\Ctvar_r}{\Ctype_\Cexp}{\Ccc_\Cexp}}{\Cff}\\
        &y\notin\Fdom{\Cenv}
    \end{align*}
    By inversion we get
	\[\truleenv{x_r:\Ctvar_r,f_r:\Frec{\Ctvar_r}{\Ctype_\Cexp}{\Ccc_\Cexp}}{\Cexp}{\Ctype_\Cexp}{\Ccc_\Cexp}\]
	By alpha-conversion we can assume that $x_r\neq y$ and
	$f_r\neq y$.  Thus, with
	\[y\notin\Fdom{\Cenv,x_r:\Ctvar_r,f_r:\Frec{\Ctvar_r}{\Ctype_\Cexp}{\Ccc_\Cexp}}\]
	we apply the induction hypothesis and get
	\[\truleenv{y:\Ctype_y,x_r:\Ctvar_r,f_r:\Frec{\Ctvar_r}{\Ctype_\Cexp}{\Ccc_\Cexp}}{\Cexp}{\Ctype_\Cexp}{\Ccc_\Cexp}\]
	As all preconditions still hold, we can apply \textsc{T-Rec} and
	finally conclude
    \[\truleenv{y:\Ctype_y}{\Frec{f_r}{x_r}{\Cexp}}{\Frec{\Ctvar_r}{\Ctype_\Cexp}{\Ccc_\Cexp}}{\Cff}\]
	\item\case{T-Pattern-Matching} By assumption we have
    \begin{align*}
        &\trule{\Fmatch{\Cexp_0}{[\Cctor_i(\bar{x})\rightarrow \Cexp_i]}}{\Ctype}{\Ccc_0 \vee \Ccc'}\\
        &y\notin\Fdom{\Cenv}
    \end{align*}
    We use inversion to deduce both
	\[\pruleenv{\Ctt;\Ctype_0}{[\Cctor_i(\bar{x})\rightarrow \Cexp_i]}{\Ctype}{\Ccc'}\]
    and
	\[\trule{\Cexp_0}{\Ctype_0}{\Ccc_0}\]
    We can directly apply the induction hypothesis
	on both judgements and obtain
	\[\pruleenvg{\Ctt;\Ctype_0}{y:\Ctype_y}{[\Cctor_i(\bar{x})\rightarrow \Cexp_i]}{\Ctype}{\Ccc'}\]
    and
	\[\truleenv{y:\Ctype_y}{\Cexp_0}{\Ctype_0}{\Ccc_0}\]
    respectively.
    Applying \textsc{T-Pattern-Matching} concludes
    \[\truleenv{y:\Ctype_y}{\Fmatch{\Cexp_0}{
	[\Cctor_i(\bar{x})\rightarrow \Cexp_i]}}{\Ctype}{\Ccc_0 \vee \Ccc'}\]
	\item\case{T-Pattern-Empty} Let $\Cenv'=\Cenv,y:\Ctype_y$ and
	\[\pruleenvown{\Ccc_0;\Ctype_0}{\Cenv'}{\Cemptypattern}{\Ctbot}{\Ccc_0}\] trivially holds by
	\textsc{T-Pattern-Empty}.
	\item\case{T-Pattern-Next} Assume
	\begin{align*}
        &\pruleenv{\Ccc_0;\Ctype_0}{[\Cctorxn\rightarrow \Cexp] \append R}{\Ctype' \cup
	\Ctype_\Cexp}{(\Ccc_0\wedge \Fhasctor{\Cctor}{\Ctype_0} \wedge \Ccc_\Cexp) \vee \Ccc'}\\
        &y\notin\Fdom{\Cenv}
    \end{align*}
    By inversion we get
    \[\pruleenv{(\Ccc_0\wedge
	\Fhasnctor{C}{\Ctype_0});\Ctype_0}{R}{\Ctype'}{\Ccc'}\] and \[\truleenv{x_i:
	\Ftget{\Ctype_0}{\Cctor}{i}}{\Cexp}{\Ctype_\Cexp}{\Ccc_\Cexp}\] for $i=1,\dots,n$. By
	alpha-conversion we know, that $\forall i=1,\dots,n.\
	x_i\neq y$. Thus, we can apply the induction hypothesis on both judgements
	and get \[\pruleenvg{(\Ccc_0\wedge \Fhasnctor{C}{\Ctype_0};\Ctype_0}{y:\Ctype_y}{R}{\Ctype'}{\Ccc'}\]
	and \[\truleenv{y:\Ctype_y,x_i:
	\Ftget{\Ctype_0}{\Cctor}{i}}{\Cexp}{\Ctype_\Cexp}{\Ccc_\Cexp}\]
    respectively.  Now we apply
	\textsc{T-Pattern-Next} and conclude
	\[\pruleenvg{\Ccc_0;\Ctype_0}{y:\Ctype_y}{[\Cctorxn\rightarrow \Cexp] \append R}{\Ctype' \cup
	\Ctype_\Cexp}{(\Ccc_0\wedge \Fhasctor{C}{\Ctype_0} \wedge \Ccc_\Cexp) \vee \Ccc'}\]
\end{description}
\end{proof}

\begin{proof}[\cref{lem:ConsistencyOfTypeSubstitution}]
Proof by induction on the derivation of the analysis.
\begin{description}
	\item\case{T-Integer}
		Let $\Cenv'=\amato{(\Cenv, y:\Ctvar)}$. By substitution and
		\textsc{T-Integer} our consequence
		\[\truleenvown{\Cenv'}{n}{\amato{n}}{\amato{\Cff}}\] immediately
		holds.
	\item\case{T-Identifier} We have two subcases.
		\begin{description}
			\item\subcase{$y=x$} We have to show that
			\[\truleenvown{\amato{(\Cenv,y:\Ctvar)}}{y}{\Ctype_\Ctvar}{\Cff}\] holds.
			We apply inversion and get
            \[(\amato{(\Cenv, y:\Ctvar)})(y)=\Ctype_\Ctvar\]
            By definition of substitution this is
			equivalent to
            \[(\amato{\Cenv},y:\Ctype_\Ctvar)(y)=\Ctype_\Ctvar\]
            As we require $\Cenv$ to be well-formed this holds trivially.

			\item\subcase{$y\neq x$} As assumption we have
			\[\truleenvown{\amato{(\Cenv, y:\Ctvar)}}{x}{\Ctvar'}{\Cff}\]
            By inversion we get $(\Cenv, y:\Ctvar)(x)=\Ctvar'$. Thus, we know
			that $\Cenv=\Cenv_1,x:\Ctvar',\Cenv_2$.  As $\Cenv$ is
			well-formed, we can insert a substitution without changing the
			equality
            \[(\amato{(\Cenv_1,x:\Ctvar',\Cenv_2,
			y:\Ctvar)})(x)=\Ctvar'\]
            Now, we can apply \textsc{T-Identifier}
			and finally get \[\truleenvown{(\amato{\Cenv_1,x:\Ctvar',\Cenv_2,
			y:\Ctvar)}}{x}{\Ctvar'}{\Cff}\]
            which is equivalent to our goal
			\[\truleenvown{\amato{(\Cenv,
			y:\Ctvar)}}{x}{\amato{\Ctvar'}}{\amato{\Cff}}\] by the rules of
			substitution.
        \end{description}
	\item\case{T-Constructor} On the assumption
	\[\truleenv{y:\Ctvar}{\Cctorenabbr}{\Ctctortnabbr}{\vee \bar{\Ccc}}\] we apply
	inversion and get \[\truleenv{y:\Ctvar}{\Cexp_i}{\Ctype_i}{\Ccc_i}\]  With the
	induction hypothesis, we can deduce
	\[\truleenvown{\amato{(\Cenv,y:\Ctvar)}}{\Cexp_i}{\amato{\Ctype_i}}{\amato{\Ccc_i}}\]
	Applying \textsc{T-Constructor} yields
	\[\truleenvown{\amato{(\Cenv,y:\Ctvar)}}{\Cctorenabbr}{\Cctor(\overline{\amato{\Ctype}})}
	{\vee(\overline{\amato{\Ccc}})}\] which is, by rules of substitution,
	equivalent to
	\[\truleenvown{\amato{(\Cenv,y:\Ctvar)}}{\Cctorenabbr}{\amato{C(\bar{\Ctype})}}
	{\amato{(\vee\bar{\Ccc})}}\]
	\item\case{T-FunApp} On the assumption
	\[\truleenv{y:\Ctvar}{\Fapp{\Cexp_1}{\Cexp_2}}{\Ftapp{\Ctype_1}{\Ctype_2}}{\Fcc{\Ctype_1}{\Ctype_2}
	\vee \Ccc_1 \vee \Ccc_2 \vee \Fhasnfun{\Ctype_1}}\] we apply inversion and get
	\[\truleenv{y:\Ctvar}{\Cexp_i}{\Ctype_i}{\Ccc_i}\]
    With the induction hypothesis, we can deduce
	\[\truleenvown{\amato{(\Cenv,y:\Ctvar)}}{\Cexp_i}{\amato{\Ctype_i}}{\amato{\Ccc_i}}\]
	Applying \textsc{T-FunApp} yields
    \begin{multline*}
        \truleenvown{\amato{(\Cenv,y:\Ctvar)}}{\Fapp{\Cexp_1}{\Cexp_2}}
        {\Ftapp{\amato{\Ctype_1}}{\amato{\Ctype_2}}\\}
        {\Fcc{\amato{\Ctype_1}}{\amato{\Ctype_2}}
        \vee \amato{\Ccc_1} \\
        \vee \amato{\Ccc_2} \vee \Fhasnfun{\amato{\Ctype_1}}}
    \end{multline*}
    which is, by rules of substitution, equivalent to
    \begin{multline*}
        \truleenvown{\amato{(\Cenv,y:\Ctvar)}}{\Fapp{\Cexp_1}{\Cexp_2}}{\amato{\Ftapp{\Ctype_1}{\Ctype_2}}\\}
        {\amato{(\Fcc{\Ctype_1}{\Ctype_2}
        \vee \Ccc_1 \vee \Ccc_2 \vee \Fhasnfun{\Ctype_1})}}
    \end{multline*}    
	\item\case{T-Rec} We assume
	\[\truleenv{y:\Ctvar}{\Frec{f_r}{x_r}{\Cexp}}{\Ftfun{\Ctvar_r}{\Ctype_\Cexp}{\Ccc_\Cexp}}{\Cff}\]
	By inversion we can follow that $y\neq f_r$, $y\neq x_r$, and
	$\Ctvar\neq\Ctvar_r$.  Additionally,
	\[\truleenv{y:\Ctvar,f_r:\Ftfun{\Ctvar_r}{\Ctype_\Cexp}{\Ccc_\Cexp},x_r:\Ctvar_r}{\Cexp}{\Ctype_\Cexp}{\Ccc_\Cexp}\]
	With the induction hypothesis we get
	\[\truleenvown{\amato{(\Cenv,y:\Ctvar,f_r:\Ftfun{\Ctvar_r}{\Ctype_\Cexp}{\Ccc_\Cexp},x_r:\Ctvar_r)}}{\Cexp}{\amato{\Ctype_\Cexp}}{\Ccc_\Cexp[\Ctype\mapsto
	\Ctype_\Ctvar]}\] 
    When we apply \textsc{T-Rec} and exploit substitution we
	finally obtain
	\[\truleenvown{\amato{(\Cenv,y:\Ctvar)}}{\Frec{f_r}{x_r}{\Cexp}}{\amato{(\Ftfun{\Ctvar_r}{\Ctype_\Cexp}{\Ccc_\Cexp})}}{\Cff}\]
	\item\case{T-Pattern-Matching} We have the assumption
	\[\truleenv{y:\Ctvar}{\Fmatch{\Cexp_0}{[\Cctor_i(\bar{x})\rightarrow
	\Cexp_i]}}{\Ctype}{\Ccc_0 \vee \Ccc'}\]
    By inversion we get two judgements. On the
	first, $\truleenv{y:\Ctvar}{\Cexp_0}{\Ctype_0}{\Ccc_0}$, we apply the induction
	hypothesis and get
	\[\truleenvown{\amato{(\Cenv,y:\Ctvar)}}{\Cexp_0}{\amato{\Ctype_0}}{\amato{\Ccc_0}}\]
	On the second judgment
	\[\pruleenvown{\Ctt;\Ctype_0}{\Cenv,y:\Ctvar}{[\Cctor_i(\bar{x})\rightarrow
	\Cexp_i]}{\Ctype}{\Ccc'}\] we apply the induction hypothesis, too and
	obtain
	\begin{multline*}
    \pruleenvown{\amato{\Ctt};\amato{\Ctype_0}}{\amato{(\Cenv,y:\Ctvar)}}{[\Cctor_i(\bar{x})\rightarrow
        \Cexp_i]\\}
        {\amato{\Ctype}}{\amato{\Ccc'}}
    \end{multline*}
    Now we can apply
	\textsc{T-Pattern-Matching} and conclude
	\begin{multline*}
    \truleenvown{\amato{(\Cenv,y:\Ctvar)}}{\Fmatch{\Cexp_0}
        {[\Cctor_i(\bar{x})\rightarrow \Cexp_i]}\\}
        {\amato{\Ctype}}{\amato{(\Ccc_0 \vee \Ccc')}}
    \end{multline*}

	\item\case{T-Pattern-Empty} The consequent
	\begin{multline*}
    \pruleenvown
            {\amato{\Ccc_0};\amato{\Ctype_0}}
            {\amato{(\Cenv,x:\Ctvar)}}{\Cemptypattern\\}
            {\amato{\Ctbot}}{\amato{\Ccc_0}}
    \end{multline*}
	holds by \textsc{T-Pattern-Empty}.
	\item\case{T-Pattern-Next} As assumption we have
    \begin{multline*}
    \pruleenvown{\Ccc_0;\Ctype_0}{\Cenv,y:\Ctvar}{[\Cctorxn\rightarrow
    ] \append \Cpatterns\\}
    {\Ctype' \cup \Ctype_\Cexp}{(\Ccc_0\wedge \Fhasctor{\Cctor}{\Ctype_0} \wedge \Ccc_\Cexp) \vee
    \Ccc'}
    \end{multline*}
    By inversion we get two judgments. As first we get
	\[\truleenvown{\Cenv,y:\Ctvar,x_i: \Ftget{\Ctype_0}{\Cctor}{i}}{\Cexp}{\Ctype_\Cexp}{\Ccc_\Cexp}\] for
	$i=1,\dots,n$ with $\forall i: y\neq x_i$ by
	alpha conversion.
    Now we can apply the induction hypothesis and obtain
	\[\truleenvown{\amato{(\Cenv,y:\Ctvar,x_i:
	\Ftget{\Ctype_0}{\Cctor}{i})}}{\Cexp}{\amato{\Ctype_\Cexp}}{\amato{\Ccc_\Cexp}}\]
    for $i=1,\dots,n$.  On the
	second judgment
    \[\pruleenvown{(\Ccc_0\wedge \Fhasnctor{\Cctor}{\Ctype_0});\Ctype_0}
    {\Cenv,y:\Ctvar}{\Cpatterns}{\Ctype'}{\Ccc'}\]
    we apply the induction
	hypothesis and deduce
    \begin{multline*}
        \pruleenvown{\amato{(\Ccc_0\wedge
        \Fhasnctor{\Cctor}{\Ctype_0})};\amato{\Ctype_0}}{\amato{(\Cenv,y:\Ctvar)}}{\Cpatterns\\}
        {\amato{\Ctype'}}{\amato{\Ccc'}}
    \end{multline*}
	Finally, we apply \textsc{T-Pattern-Next} and conclude that
	\begin{multline*}
    \pruleenvown{\amato{\Ccc_0};\amato{\Ctype_0}}{\Cenv,y:\Ctvar}{[\Cctorxn\rightarrow
	\Cexp] \append \Cpatterns\\}
    {\amato{(\Ctype' \cup \Ctype_\Cexp)}}{\amato{((\Ccc_0\wedge \Fhasctor{\Ctype_0}
	\wedge \Ccc_\Cexp) \vee \Ccc')}}
    \end{multline*}
    holds.
\end{description}
\end{proof}

\begin{proof}[\cref{lem:ConsistencyOfValueSubstitution}]
Proof by induction on the derivation of the analysis.
\begin{description}
	\item\case{T-Integer} The consequent $\trule{\xmato{n}}{n}{\Cff}$
	trivially holds by \textsc{T-Integer}.
	\item\case{T-Identifier} For this case we have two subcases.
	One with $y=x$ and one with $y\neq x$.
	\begin{description}
		\item\subcase{$y=x$} We have the assumptions
        \begin{align*}
            &\truleenv{y:\Ctype_y}{y}{\Ctype}{\Cff}\\
            &\trule{v}{\Ctype_y}{\Cff}
        \end{align*}
        By inversion we get
		$(\Cenv,y:\Ctype_y)(y)=\Ctype$ and thus, $\Ctype=\Ctype_y$ and $\trule{v}{\Ctype}{\Cff}$.
        By applying substitution on our goal
        \[\trule{\xmato{y}}{\Ctype}{\Cff}\] we get
		$\trule{v}{\Ctype}{\Cff}$ and are done.
		\item\subcase{$y\neq x$} We have the assumptions
        \begin{align*}
            &\truleenv{y:\Ctype_y}{x}{t}{\Cff}\\
            &\trule{v}{\Ctype_y}{\Cff}
        \end{align*}
        By inversion we get $(\Cenv,y:\Ctype_y)(x)=t$. Thus, $\Cenv=\Cenv_1,x:t,\Cenv_2$ which
		leads to $(\Cenv_1,x:t,\Cenv_2)(x)=t$.  This is a precondition
		for $\truleenv{x:t}{x}{t}{\Cff}$ which by rules of substitution is
		equivalent to $\truleenv{x:t}{\xmato{x}}{t}{\Cff}$.
	\end{description}
	\item\case{T-Constructor} Assume
	\[\truleenv{y:\Ctype_y}{\Cctorenabbr}{\Ctctortnabbr}{\vee\bar{\Ccc}}\]
    By inversion
	we get \[\truleenv{y:\Ctype_y}{\Cexp_i}{\Ctype_i}{\Ccc_i}\] for $i=1,\dots,n$
    and apply the induction
	hypothesis.  Thus, we obtain 
    \[\trule{\xmato{\Cexp_i}}{\Ctype_i}{\Ccc_i}\]
    and can
	apply \textsc{T-Constructor} and conclude
	\[\trule{C(\overline{\xmato{\Cexp}})}{\Ctctortnabbr}{\vee\bar{\Ccc}}\]
    which is by
	substitution equivalent to
	\[\trule{\xmato{\Cctorenabbr}}{\Ctctortnabbr}{\vee\bar{\Ccc}}\]
	\item\case{T-FunApp} Assume
	\[\truleenv{y:\Ctype_y}{\Fapp{\Cexp_1}{\Cexp_2}}{\Ftapp{\Ctype_1}{\Ctype_2}}{\Fcc{\Ctype_1}{\Ctype_2} \vee
	\Ccc_1 \vee \Ccc_2 \vee \Fhasnfun{\Ctype_1}}\]
    By inversion we get
	\[\truleenv{y:\Ctype_y}{\Cexp_i}{\Ctype_i}{\Ccc_i}\]
    and apply the induction hypothesis.
	Thus, we obtain \[\trule{\xmato{\Cexp_i}}{\Ctype_i}{\Ccc_i}\] and can apply
	\textsc{T-FunApp} and conclude
	\[\trule{\Fapp{\xmato{\Cexp_1}}{\xmato{\Cexp_2}}}{\Ftapp{\Ctype_1}{\Ctype_2}}{\Fcc{\Ctype_1}{\Ctype_2}
	\vee \Ccc_1 \vee \Ccc_2 \vee \Fhasnfun{\Ctype_1}}\]
    which is equivalent, by the
	rules of substitution, to
	\[\trule{\xmato{(\Fapp{\Cexp_1}{\Cexp_2})}}{\Ftapp{\Ctype_1}{\Ctype_2}}{\Fcc{\Ctype_1}{\Ctype_2} \vee
	\Ccc_1 \vee \Ccc_2 \vee \Fhasnfun{\Ctype_1}}\]
	\item\case{T-Rec} As part of the antecedent we have $\truleenv{y:
	\Ctype_y}{\Frec{f_r}{x_r}{e}}{\Frec{\Ctvar_r}{\Ctype_e}{\Ccc_e}}{\Cff}$. By
	inversion we get
	$\truleenv{y:\Ctype_y,f_r:\Frec{\Ctvar_r}{\Ctype_e}{\Ccc_e},x_r:\Ctvar_r}{e}{\Ctype_e}{\Ccc_e}$.
	Our induction hypothesis contains the antecedent $\trule{v}{\Ctype_y}{\Cff}$ but
	we need $\truleenv{f_r:\Frec{\Ctvar_r}{\Ctype_e}{\Ccc_e}}{v}{\Ctype_y}{\Cff}$ which
	we get using weakening (\cref{lem:Weakening}). Now we can
	apply the induction hypothesis and obtain
	$\truleenv{f_r:\Frec{\Ctvar_r}{\Ctype_e}{\Ccc_e},x_r:\Ctvar_r}{\xmato{e}}{\Ctype_e}{\Ccc_e}$.
	Applying \textsc{T-Rec} and the rules of substitution finally yields
	$\trule{\xmato{(\Frec{f_r}{x_r}{e})}}{\Frec{\Ctvar_r}{\Ctype_e}{\Ccc_e}}{\Cff}$.
	\item\case{T-Pattern-Matching} We have $\truleenv{y: \Ctype_y}{\Fmatch{\Cexp_0}
	{[\Cctorxnabbr\rightarrow e_i]}}{t}{\Ccc_0 \vee \Ccc'}$. By inversion
	we obtain $\pruleenvg{\Ctt;\Ctype_0}{y:\Ctype_y}{[\Cctorxnabbr\rightarrow
	e_i]}{t}{\Ccc'}$ and $\truleenv{y: \Ctype_y}{\Cexp_0}{\Ctype_0}{\Ccc_0}$. We apply the
	induction hypothesis on both preconditions and get
	$\pruleenv{\Ctt;\Ctype_0}{\xmato{\{C_i\bar{x}\rightarrow e_i\}}}{t}{\Ccc'}$ and
	$\trule{\xmato{\Cexp_0}}{\Ctype_0}{\Ccc_0}$ which are the preconditions for
	\textsc{T-Pattern-Matching}.  We conclude by applying the rules of
	substitution and get $\trule{\xmato{(\Fmatch{e_0}
	{[\Cctorxnabbr\rightarrow e_i]})}}{t}{\Ccc_0 \vee \Ccc'}$.
	\item\case{T-Pattern-Empty} Our goal, $\pruleenv{\Ccc_0;\Ctype_0}{\xmato{\{
	\}}}{\Ctbot}{\Ccc_0}$, trivially holds by substitution and \textsc{T-Pattern-Empty}.
	\item\case{T-Pattern-Next} We assume
	$\pruleenvg{\Ccc_0;\Ctype_0}{y:\Ctype_y}{\{C(\bar{x}) \rightarrow
	e\};\bar{p}}{t' \cup \Ctype_e}{(\Ccc_0\wedge \Fhasctor{C}{\Ctype_0} \wedge \Ccc_e) \vee
	\Ccc'}$. By inversion we get (1) $\pruleenvg{(\Ccc_0\wedge \Fhasnctor{C}{\Ctype_0});\Ctype_0}{y:\Ctype_y}{\bar{p}}{t'}{\Ccc'}$ and (2) $\truleenv{y:\Ctype_y, x_i: \Ftget{\Ctype_0}{C}{i}}{e}{\Ctype_e}{\Ccc_e}$ with $i=1,\dots,n$.
	On (1) we can apply the induction
	hypothesis and get $\pruleenv{(\Ccc_0\wedge C.n\nsubseteq
	\Ctype_0);\Ctype_0}{\xmato{\bar{p}}}{t'}{\Ccc'}$. On (2) the antecedent of the 
	induction hypothesis does not fit $\trule{v}{\Ctype_y}{\Cff}$. Thus we apply
	weakening (\cref{lem:Weakening}) and obtain $\truleenv{x_i: \Ftget{\Ctype_0}{C}{i}}{v}{\Ctype_y}{\Cff}$ with
	$i=1,\dots,n$. Now we can apply the induction hypothesis and get $\truleenv{x_i: \Ftget{\Ctype_0}{C}{i}}{\xmato{e}}{\Ctype_e}{\Ccc_e}$.
	Using the rule \textsc{T-Pattern-Next} and substitution,
	we conclude that $\pruleenv{\Ccc_0;\Ctype_0}{\xmato{(\{C(\bar{x}) \rightarrow
	e\};\bar{p})}}{t' \cup \Ctype_e}{(\Ccc_0\wedge \Fhasctor{C}{\Ctype_0} \wedge \Ccc_e) \vee
	\Ccc'}$ holds.
\end{description}
\end{proof}

\begin{proof}[\cref{lem:UnsatisfiabilityAfterMatchingPatterns}]
Proof by induction on the derivation of $\vdash_p$.
\begin{description}
	\item\case{T-Pattern-Empty} For
    \[\pruleenv{\Cff;\Ctype_0}{\Cemptypattern}{\Ctbot}{\Cff}\]
		we have $\nvDash \Cff$.
	\item\case{T-Pattern-Next} We have
    \[\pruleenv{\Ccc_0;\Ctype_0}{[\Cctorxn\rightarrow \Cexp]\append\Cpatterns}{\Ctype' \cup \Ctype_\Cexp}{(\Ccc_0\wedge \Fhasctor{\Cctor}{\Ctype_0} \wedge \Ccc_\Cexp)
    \vee \Ccc'}\]
    and know that $\Ccc_0=\Cff$. Thus, the crash condition
	simplifies to $(\Cff \wedge \Fhasctor{\Ctype_0} \wedge \Ccc_\Cexp) \vee \Ccc'=\Ccc'$.
	We have to show that $\nvDash \Ccc'$.
	By inversion we get
    \[\pruleenv{(\Ccc_0\wedge \Fhasnctor{\Cctor}{\Ctype_0});\Ctype_0}{\Cpatterns}{\Ctype'}{\Ccc'}\]
	and with $\Ccc_0=\Cff$ we get
    \[\pruleenv{\Cff;\Ctype_0}{\Cpatterns}{\Ctype'}{\Ccc'}\]
	Applying the induction hypothesis finally yields $\nvDash \Ccc'$.
\end{description}
\end{proof}

\begin{proof}[\cref{thm:Preservation}]
We abbreviate $\mathcal{E}[\hat{e}]$ with $e$ and $\mathcal{E}[\hat{e}']$ with
$e'$, respectively.  By induction on the derivation of $e\hookrightarrow e'$.
In all the cases building on \textsc{Eval-Hole}, we have
$\hat{e}\hookrightarrow\hat{e}'$ by inversion on $e\hookrightarrow e'$.
\begin{description}
	\item\case{$e=C(v_1\dots,v_n,\hat{e},e_1,\dots,e_m) \hookrightarrow
	e'=C(v_1\dots,v_n,\hat{e}',e_1,\dots,e_m)$}   By assumption we have
	\begin{alignat}{11}
		&\trule{C(v_{1\dots n},\hat{e},e_{1\dots m})&&}{C(\Ctype_{v_{1\dots
		n}},\hat{\Ctype},\Ctype_{\Cexp_{1\dots m}})&&}{\vee_{i=1}^{n} \Ccc_{v_i} \vee
		\hat{\Ccc} \vee \vee_{i=1}^m
		\Ccc_{\Cexp_i}}\label{equ:them_pres_ctor_one}\\
		&\trule{C(v_{1\dots n},\hat{e}',e_{1\dots m})&&}{C(\Ctype_{v_{1\dots
		n}}',\hat{\Ctype}',\Ctype_{\Cexp_{1\dots m}}')&&}{\vee_{i=1}^{n} \Ccc_{v_i}' \vee
		\hat{\Ccc}' \vee \vee_{i=1}^m
		\Ccc_{\Cexp_i}'}\label{equ:them_pres_ctor_two}
	\end{alignat}
	Inversion on equation~\eqref{equ:them_pres_ctor_one} yields
	\[\trule{v_i}{\Ctype_{v_i}}{\Ccc_{v_i}} \qquad \trule{\Cexp_i}{\Ctype_{\Cexp_i}}{\Ccc_{\Cexp_i}}
	\qquad \trule{\hat{e}}{\hat{\Ctype}}{\hat{\Ccc}}\] and for inversion on
	equation~\eqref{equ:them_pres_ctor_two} we get
	\[\trule{v_i}{\Ctype_{v_i}'}{\Ccc_{v_i}'} \qquad
	\trule{\Cexp_i}{\Ctype_{\Cexp_i}'}{\Ccc_{\Cexp_i}'} \qquad
	\trule{\hat{e}'}{\hat{\Ctype}'}{\hat{\Ccc}'}\] Thus, we can follow that
	$\Ctype_{v_i}=\Ctype_{v_i}'$, $\Ctype_{\Cexp_i}=\Ctype_{\Cexp_i}'$, $\Ccc_{v_i}=\Ccc_{v_i}'$, and
	$\Ccc_{\Cexp_i}=\Ccc_{\Cexp_i}'$. Additionally, we have $(\hat{\Ctype}=\hat{\Ctype}' \wedge
	\hat{\Ccc}\leftrightarrow\hat{\Ccc}') \rightarrow t=t' \wedge \Ccc=\Ccc'$.
	Now we can apply the induction hypothesis
	($\trule{\hat{e}}{\hat{\Ctype}}{\hat{\Ccc}} \wedge \hat{e} \hookrightarrow
	\hat{e'} \wedge \trule{\hat{e}'}{\hat{\Ctype}'}{\hat{\Ccc}'} \rightarrow
	\hat{\Ctype}=\hat{\Ctype}' \wedge \hat{\Ccc} \leftrightarrow \hat{\Ccc}'$) and thus
	conclude that $t=t'$ and $\Ccc\leftrightarrow\Ccc'$.
	\item\case{$e=\Fapp{\hat{e}}{\Cexp_2} \hookrightarrow e'=\Fapp{\hat{e}'}{\Cexp_2}$}
	By assumption we have
	\begin{alignat}{11}
		&\trule{\Fapp{\hat{e}}{\Cexp_2}&&}{\Ftapp{\hat{\Ctype}}{\Ctype_2}&&}{\Fcc{\hat{\Ctype}}{\Ctype_2}
		\vee \hat{\Ccc} \vee \Ccc_2 \vee
		\Fhasnfun{\hat{\Ctype}}}\label{equ:them_pres_ee_one}\\
		&\trule{\Fapp{\hat{e}'}{\Cexp_2}&&}{\Ftapp{\hat{\Ctype}'}{\Ctype_2'}&&}{\Fcc{\hat{\Ctype}'}{\Ctype_2'}
		\vee \hat{\Ccc}' \vee \Ccc_2' \vee
		\Fhasnfun{\hat{\Ctype}'}}\label{equ:them_pres_ee_two}
	\end{alignat}
	When applying inversion on equation~\eqref{equ:them_pres_ee_one} we obtain
	\[\trule{\hat{e}}{\hat{\Ctype}}{\hat{\Ccc}} \qquad \trule{\Cexp_2}{\Ctype_2}{\Ccc_2}\]
	and for inversion on equation~\eqref{equ:them_pres_ee_two} we get
	\[\trule{\hat{e}'}{\hat{\Ctype}'}{\hat{\Ccc}'} \qquad
	\trule{\Cexp_2}{\Ctype_2'}{\Ccc_2'}\] Thus, we have $\Ctype_2=\Ctype_2'$ and $\Ccc_2=\Ccc_2'$
	and $(\hat{\Ctype}=\hat{\Ctype}' \wedge \hat{\Ccc}\leftrightarrow\hat{\Ccc}')
	\rightarrow t=t' \wedge \Ccc=\Ccc'$.  We apply the induction hypothesis
	($\trule{\hat{e}}{\hat{\Ctype}}{\hat{\Ccc}} \wedge \hat{e} \hookrightarrow
	\hat{e'} \wedge \trule{\hat{e}'}{\hat{\Ctype}'}{\hat{\Ccc}'} \rightarrow
	\hat{\Ctype}=\hat{\Ctype}' \wedge \hat{\Ccc} \leftrightarrow \hat{\Ccc}'$) and
	conclude that $t=t'$ and $\Ccc\leftrightarrow\Ccc'$.
	\item\case{$e=\Fapp{v}{\hat{e}} \hookrightarrow e'=\Fapp{v}{\hat{e}'}$}
	By assumption we have 
	\begin{alignat}{11}
		&\trule{\Fapp{v}{\hat{e}}&&}{\Ftapp{\Ctype_1}{\hat{\Ctype}}&&}{\Fcc{\Ctype_1}{\hat{\Ctype}}
		\vee \Ccc_1 \vee \hat{\Ccc} \vee
		\Fhasnfun{\Ctype_1}}\label{equ:them_pres_ev_one}\\
		&\trule{\Fapp{v}{\hat{e}'}&&}{\Ftapp{\Ctype_1}{\hat{\Ctype}'}&&}{\Fcc{\Ctype_1}{\hat{\Ctype}'}
		\vee \Ccc_1 \vee \hat{\Ccc}' \vee
		\Fhasnfun{\Ctype_1}}
		\label{equ:them_pres_ev_two}
	\end{alignat}
	Inversion on equation~\eqref{equ:them_pres_ev_one} yields
	\[\trule{\hat{e}}{\hat{\Ctype}}{\hat{\Ccc}} \qquad \trule{v}{\Ctype_1}{\Ccc_1}\]
	and inversion on equation~\eqref{equ:them_pres_ev_two} yields
	\[\trule{\hat{e}'}{\hat{\Ctype}'}{\hat{\Ccc}'} \qquad \trule{v}{\Ctype_1'}{\Ccc_1'}\]
	Thus we conclude $\Ctype_1=\Ctype_1'$, $\Ccc_1=\Ccc_1'$, and furthermore,
	$(\hat{\Ctype}=\hat{\Ctype}' \wedge \hat{\Ccc}\leftrightarrow\hat{\Ccc}')
	\rightarrow t=t' \wedge \Ccc=\Ccc'$.  We apply the induction hypothesis
	($\trule{\hat{e}}{\hat{\Ctype}}{\hat{\Ccc}} \wedge \hat{e} \hookrightarrow
	\hat{e'} \wedge \trule{\hat{e}'}{\hat{\Ctype}'}{\hat{\Ccc}'} \rightarrow
	\hat{\Ctype}=\hat{\Ctype}' \wedge \hat{\Ccc} \leftrightarrow \hat{\Ccc}'$) and
	conclude that $t=t'$ and $\Ccc\leftrightarrow\Ccc'$.
	\item\case{$e=\Fmatch{\hat{e}}{\auxcases}\hookrightarrow e'=\Fmatch{\hat{e}'}{\auxcases}$}
	By assumption we have
	\begin{alignat}{11}
		&\trule{\Fmatch{\hat{e}}{\auxcases}&&}{\Ctype_p&&}{\hat{\Ccc} \vee \Ccc_p}
		\label{equ:them_pres_match_one}\\
		&\trule{\Fmatch{\hat{e}'}{\auxcases}&&}{\Ctype_p'&&}{\hat{\Ccc}' \vee \Ccc_p'}
		\label{equ:them_pres_match_two}
	\end{alignat}
	By inversion on equation~\eqref{equ:them_pres_match_one} we get
	\begin{align}
		&\trule{\hat{e}}{\hat{\Ctype}}{\hat{\Ccc}}\nonumber\\
		\pruleenvown{\Ctt;\hat{\Ctype}}{&\Cenv}{\auxcases}{\Ctype_p}{\Ccc_p}\label{equ:them_pres_match_three}
	\end{align}
	and by inversion on equation~\eqref{equ:them_pres_match_two} we obtain
	\begin{align}
		&\trule{\hat{e}'}{\hat{\Ctype}'}{\hat{\Ccc}'}\nonumber\\
		\pruleenvown{\Ctt;\hat{\Ctype}'}{&\Cenv}{\auxcases}{\Ctype_p'}{\Ccc_p'}\label{equ:them_pres_match_four}
	\end{align}
	Now by applying the induction hypothesis ($\trule{\hat{e}}{\hat{\Ctype}}{\hat{\Ccc}} \wedge \hat{e} \hookrightarrow
	\hat{e'} \wedge \trule{\hat{e}'}{\hat{\Ctype}'}{\hat{\Ccc}'} \rightarrow
	\hat{\Ctype}=\hat{\Ctype}' \wedge \hat{\Ccc} \leftrightarrow \hat{\Ccc}'$)
	we can conclude that the equations~\eqref{equ:them_pres_match_three} and~\eqref{equ:them_pres_match_four}
	are equivalent and thus, $t=t'$ and $\Ccc=\Ccc'$.
\end{description}
Now to the interesting cases.
\begin{description}
	\item\case{$e=\Fapp{(\Frec{f}{x}{\hat{e}})}{v} \hookrightarrow
	e'=\auxsubst{\hat{e}}{x\mapsto v, f\mapsto \Frec{f}{x}{\hat{e}}}$}
	By assumption we have
	\begin{align}
		\trule{\Fapp{(\Frec{f}{x}{\hat{e}}}{v})}{\Ftapp{\Ctype_1}{\Ctype_2}}{\Fcc{\Ctype_1}{\Ctype_2} \vee \Ccc_1 \vee \Ccc_2 \vee \Fhasnfun{\Ctype_1}}
		\label{equ:them_pres_recapp_assumption_type_one}
	\end{align}
	on which we apply inversion (\textsc{T-FunApp}) and get
	\begin{align}
		&\trule{v}{\Ctype_v}{\Cff}\nonumber\\
		&\trule{\Frec{f}{x}{\hat{e}}}{\Frec{\Ctvar}{\Ctype_e}{\Ccc_e}}{\Cff}\label{equ:them_pres_recapp_rec}
	\end{align}
	Applying inversion (\textsc{T-Rec}) on the second judgement~\eqref{equ:them_pres_recapp_rec} yields
	\begin{align}
		\truleenv{x:\Ctvar,f:\Frec{\Ctvar}{\Ctype_e}{\Ccc_e}}{\hat{e}}{\Ctype_e}{\Ccc_e}\qquad \Ctvar\ \textnormal{fresh}\label{equ:them_pres_recapp_rectop}
	\end{align}
	Thus, for equation~\eqref{equ:them_pres_recapp_assumption_type_one} we have
	$\Ftapp{\Ctype_1}{\Ctype_2} = \Ftapp{(\Frec{\Ctvar}{\Ctype_e}{\Ccc_e})}{\Ctype_v} = \auxsubst{\Ctype_e}{\Ctvar\mapsto \Ctype_v}$
	and $\Fcc{\Ctype_1}{\Ctype_2} \vee \Ccc_1 \vee \Ccc_2 \vee \Fhasnfun{\Ctype_1}
		  = \Fcc{(\Frec{\Ctvar}{\Ctype_e}{\Ccc_e})}{\Ctype_v} \vee \Cff \vee \Cff \vee \Fhasnfun{(\Frec{\Ctvar}{\Ctype_e}{\Ccc_e})}
		  = \auxsubst{\Ccc_e}{\Ctvar\mapsto \Ctype_v}$.
	Our assumption can thus be rewritten as
	\begin{align*}
		\trule{\Fapp{(\Frec{f}{x}{\hat{e}}}{v})}{\auxsubst{\Ctype_e}{\Ctvar\mapsto \Ctype_v}}{\auxsubst{\Ccc_e}{\Ctvar\mapsto \Ctype_v}}
	\end{align*}
	We now claim that the following holds
	\begin{align*}
		\trule{\auxsubst{\hat{e}}{x\mapsto v, f\mapsto \Frec{f}{x}{\hat{e}}}}{\auxsubst{\Ctype_e}{\Ctvar\mapsto \Ctype_v}}{\auxsubst{\Ccc_e}{\Ctvar\mapsto \Ctype_v}}
	\end{align*}
	To show that, we apply \cref{lem:ConsistencyOfValueSubstitution} and get
	\begin{align*}
		&\truleenv{x:\Ctvar,f:\Frec{\Ctvar}{\Ctype_e}{\Ccc_e}}{\hat{e}}{\Ctype_e}{\Ccc_e}\ \wedge\ 
		\trule{\Frec{f}{x}{\hat{e}}}{\Frec{\Ctvar}{\Ctype_e}{\Ccc_e}}{\Cff} \rightarrow\\
		&\quad \truleenv{x:\Ctvar}{\auxsubst{\hat{e}}{f\mapsto \Frec{f}{x}{\hat{e}}}}{\Ctype_e}{\Ccc_e}
	\end{align*}
	On the consequent, we apply \cref{lem:ConsistencyOfTypeSubstitution} and get
	\begin{align*}
		&\truleenv{x:\Ctvar}{\auxsubst{\hat{e}}{f\mapsto
		\Frec{f}{x}{\hat{e}}}}{\Ctype_e}{\Ccc_e} \rightarrow\\ &\quad
		\truleenvown{\auxsubst{\Cenv}{\Ctvar\mapsto
		\Ctype_v},x:\Ctype_v}{\auxsubst{\hat{e}}{f\mapsto
		\Frec{f}{x}{\hat{e}}}}{\auxsubst{\Ctype_e}{x\Ctvar\mapsto
		\Ctype_v}}{\auxsubst{\Ccc_e}{\Ctvar\mapsto \Ctype_v}}
	\end{align*}
	From equation~\eqref{equ:them_pres_recapp_rectop} we know that $\Ctvar$ was fresh,
	and thus $\Ctvar\notin ran(\Cenv)$ which implies $\auxsubst{\Cenv}{\Ctvar\mapsto \Ctype_v}=\Cenv$.
	Finally, we again apply \cref{lem:ConsistencyOfValueSubstitution} and get
	\begin{align*}
		&\truleenvown{\auxsubst{\Cenv}{\Ctvar\mapsto
		\Ctype_v},x:\Ctype_v}{\auxsubst{\hat{e}}{f\mapsto
		\Frec{f}{x}{\hat{e}}}}{\auxsubst{\Ctype_e}{x\Ctvar\mapsto
		\Ctype_v}}{\auxsubst{\Ccc_e}{\Ctvar\mapsto \Ctype_v}}\\
		&\quad \wedge\  \trule{v}{\Ctype_v}{\Cff} \rightarrow\\
		&\quad \trule{\auxsubst{(\auxsubst{\hat{e}}{f\mapsto
		\Frec{f}{x}{\hat{e}}})}{x\mapsto v}}{\auxsubst{\Ctype_e}{\Ctvar\mapsto
		\Ctype_v}}{\auxsubst{\Ccc_e}{\Ctvar\mapsto \Ctype_v}}
	\end{align*}
	As $x\notin free(\Frec{f}{x}{\hat{e}})$, we can reorder the substitution
	and
	\[\trule{\auxsubst{\hat{e}}{x\mapsto v, f\mapsto
	\Frec{f}{x}{\hat{e}}}}{\auxsubst{\Ctype_e}{\Ctvar\mapsto
	\Ctype_v}}{\auxsubst{\Ccc_e}{\Ctvar\mapsto \Ctype_v}} \]
	holds, as claimed.
	\item\case{$e=\Fmatch{\Cctorvnabbr}{\Cctorxnabbr \rightarrow \hat{e} | r}
	\hookrightarrow e'=\auxsubst{\hat{e}}{x_i\mapsto v_i}$ ($i=1\dots,n$)}
	By our assumptions we know that
	\begin{align}
		&\trule{\Fmatch{\Cctorvnabbr}{\Cctorxnabbr \rightarrow \hat{e} | r}}{\Ctype_p}{\Ccc_0 \vee \Ccc_p}\label{equ:them_pres_match_assume_one}\\
		&\trule{\auxsubst{\hat{e}}{x_i\mapsto v_i}}{t'}{\Ccc'}\label{equ:them_pres_match_assume_two}
	\end{align}
	Now we have to show that $\Ctype_p=t'$ and $\Ccc_0 \vee \Ccc_p \leftrightarrow \Ccc'$ holds.
	By inversion on equation~\eqref{equ:them_pres_match_assume_one} we get two judgements:
	At first, $\trule{C(\bar{v})}{C(\overline{\Ctype_v})}{\Cff}$ which by inversion yields
	\begin{align}
		\trule{v_i}{\Ctype_i}{\Cff} \qquad i=1,\dots,n
		\label{equ:them_pres_match_vi}
	\end{align}
	and second, we get
	\begin{align}
		\pruleenv{\Ctt;C(\overline{\Ctype_v})}{C(\bar{x})\rightarrow \hat{e}|r}{\Ctype_e \cup \tilde{\Ctype}}{(\Ccc_0 \wedge \Fhasnctor{C}{C(\overline{\Ctype_v}))}\wedge \Ccc_e) \vee \tilde{\Ccc}}
		\label{equ:them_pres_match_assume_inv_one}
	\end{align}
	We again apply inversion and by \textsc{T-Pattern-Next} get
	\begin{align}
		&\pruleenv{(\Ccc_0\wedge\Fhasnctor{C}{C(\overline{\Ctype_v})});C(\overline{\Ctype_v})}{r}{\tilde{\Ctype}}{\tilde{\Ccc}}\label{equ:them_pres_match_assume_inv_inv_one}\\
		&\truleenv{x_i:\Ftget{C(\overline{\Ctype_v})}{C}{i}}{\hat{e}}{\Ctype_e}{\Ccc_e}\qquad i=1,\dots,n\label{equ:them_pres_match_assume_inv_inv_two}
	\end{align}
	As we know that $\Ccc_0=\Ctt$, $\Fhasctor{C}{C(\overline{\Ctype_v})}=\Ctt$, and
	$\Fhasnctor{C}{C(\overline{\Ctype_v})}=\Cff$ (the pattern matches), we can change equation~\eqref{equ:them_pres_match_assume_inv_inv_one} to
	\begin{align*}
		\pruleenv{\Cff;C(\overline{\Ctype_v})}{r}{\tilde{\Ctype}}{\tilde{\Ccc}}
	\end{align*}
	and then apply \cref{lem:UnsatisfiabilityAfterMatchingPatterns} and conclude $\nvDash{\tilde{\Ccc}}$.
	Thus, equation~\eqref{equ:them_pres_match_assume_inv_one} simplifies to
	\begin{align*}
		\pruleenv{\Ctt;C(\overline{\Ctype_v})}{C(\bar{x})\rightarrow \hat{e}|r}{\Ctype_e \cup \tilde{\Ctype}}{\Ccc_e}
	\end{align*}
	Now we have to derive the types of equation~\eqref{equ:them_pres_match_assume_two} and show
	that $\Ctype_e \cup \tilde{\Ctype}=t'$ and $\Ccc_e\leftrightarrow \Ccc'$.
	In equation~\eqref{equ:them_pres_match_assume_inv_inv_two} we can simplify $\Ftget{C(\overline{\Ctype_v})}{C}{i}$ to $\Ctype_i$ and then apply
	\cref{lem:ConsistencyOfValueSubstitution} (value substitution) $n$ times using equation~\eqref{equ:them_pres_match_vi} with
	the appropriate $i$. Thus, we obtain
	\begin{align*}
		\trule{\auxsubst{\auxsubst{\hat{e}}{x_n\mapsto v_n}\dots}{x_1\mapsto v_1}}{\Ctype_e}{\Ccc_e}
	\end{align*}
	The substitution is order-independent as $\forall i,j: x_i\notin v_i$ and thus we get
	\begin{align*}
		\trule{\auxsubst{\hat{e}}{x_i\mapsto v_i}}{\Ctype_e}{\Ccc_e}
	\end{align*} By subtyping ($\subtype{\Ctype_e}{\Ctype_e \cup \tilde{\Ctype}}$) our claim holds.
	\item\case{$e=\Fmatch{\Cctorvnabbr}{\ectordxbar \rightarrow \hat{e} | r}
	\hookrightarrow e'=\Fmatch{\Cctorvnabbr}{r}$}
	By assumption we have
	\begin{align}
		&\trule{\Fmatch{\Cctorvnabbr}{\ectordxbar \rightarrow \hat{e} | r}}{\Ctype_p}{\Ccc_0 \vee \Ccc_p}\label{equ:them_pres_nomatch_assume_one}\\
		&\trule{\Fmatch{\Cctorvnabbr}{r}}{\Ctype_p'}{\Ccc_0' \vee \Ccc_p'}\label{equ:them_pres_nomatch_assume_two}
	\end{align}
	Our claim is that $\subtype{\Ctype_p'}{\Ctype_p}$ and $\Ccc_0 \vee \Ccc_p \leftrightarrow \Ccc_0' \vee \Ccc_p'$.
	To prove this, we apply inversion on equation~\eqref{equ:them_pres_nomatch_assume_one} and obtain
	\begin{align}
		&\trule{\Cctorvnabbr}{\Ctctortnabbr}{\Cff}\label{equ:them_pres_nomatch_val}\\
		&\pruleenv{\Ctt;\Ctctortnabbr}{\ectordxbar \rightarrow
		\hat{e}|r}{\Ctype_{\hat{e}} \cup \tilde{\Ctype}}{\Ctt \wedge \Fhasctor{C}{D} \wedge
		\Ccc_{\hat{e}} \vee \tilde{\Ccc}}\label{equ:them_pres_nomatch_assume_one_inversion}
	\end{align}
	where the crash condition ($\Ccc_0=\Cff$) of
	equation~\eqref{equ:them_pres_nomatch_val} can be deduced by another
	inversion.  Equation~\eqref{equ:them_pres_nomatch_assume_one_inversion} can
	be simplified as we know that $\Fhasctor{C}{D}=\Cff$. Thus, we get
	$\Ccc=\Ccc_p$ and (without simplifications) $t=\Ctype_p$.
	Inversion on the other assumption~\eqref{equ:them_pres_nomatch_assume_two} again yields
	equation~\eqref{equ:them_pres_nomatch_assume_two} and additionally we have
	\begin{align}
		\pruleenv{\Ctt;\Ctctortnabbr}{r}{\Ctype_p'}{\Ccc_p'}\label{equ:them_pres_nomatch_assume_two_inversion}
	\end{align}
	With the same reason as above, we can follow that $\Ccc_0=\Ccc_0'=\Cff$ and
	thus $\Ccc'=\Ccc_p'$.  Inversion on both
	equations~\eqref{equ:them_pres_nomatch_assume_one_inversion} and
	\eqref{equ:them_pres_nomatch_assume_two_inversion} requires a case
	distinction over $r$ as both the rules \textsc{T-Pattern-Empty} and
	\textsc{T-Pattern-Next} match.
	\begin{description}
		\item\subcase{$r=\Cemptypattern$}
		Inversion on equation~\eqref{equ:them_pres_nomatch_assume_one_inversion} yields
		\begin{align}
			\pruleenv{\Ctt\wedge(\phasdtor{\Cctorvnabbr} \vee \phasndtor{\Ctctortnabbr});\Ctctortnabbr}{
            \Cemptypattern}{\tilde{\Ctype}}{\tilde{\Ccc}}\nonumber
		\end{align}
		By \textsc{T-Pattern-Empty} we get $\tilde{\Ctype}=\Ctbot$ and $\tilde{\Ccc}=\Ctt$.
		Thus, $t=\Ctype_{\hat{e}}\cup \hat{\Ctype}$ and $\Ccc=\tilde{\Ccc}=\Ctt$.
		With $r=\Cemptypattern$ and \textsc{T-Pattern-Empty} equation~\eqref{equ:them_pres_nomatch_assume_two_inversion}
		gets more specific: $\pruleenv{\Ctt;\Ctctortnabbr}{\Cemptypattern}{\Ctbot}{\Ctt}$.
		It follows that $\Ctype_p'=\Ctbot$ and $\Ccc'=\Ctt$ and thus, our claim holds.
		\item\subcase{$r\neq\Cemptypattern$}
		Inversion on equation~\eqref{equ:them_pres_nomatch_assume_one_inversion} now gives
		\begin{align}
			\pruleenv{\Ctt\wedge(\phasdtor{\Cctorvnabbr} \vee \phasndtor{\Ctctortnabbr});\Ctctortnabbr}{r}{\tilde{\Ctype}}{\tilde{\Ccc}}\nonumber
		\end{align}
		This can be simplified by $\phasndtor{\Ctctortnabbr}=\Ctt$ to
		\begin{align}
			\pruleenv{\Ctt;\Ctctortnabbr}{r}{\tilde{\Ctype}}{\tilde{\Ccc}}\label{equ:them_pres_nomatch_simple_two}
		\end{align}
		As we have deterministic rules, we can follow from the equivalency of
		equation~\eqref{equ:them_pres_nomatch_assume_two_inversion} and
		\eqref{equ:them_pres_nomatch_simple_two} that $\tilde{\Ctype}=\Ctype_p'$ and
		$\tilde{\Ccc}=\Ccc_p'$. Thus, $\Ccc=\Ccc'$ and $t=t'$.
	\end{description}
\end{description}
And finally, the error cases.
\begin{description}
	\item\case{$e=\Fapp{\hat{v}}{v} \hookrightarrow e'=\Cerr$}
	By assumption we have 
	\begin{align}
		&\trule{\Fapp{\hat{v}}{v}}{t}{\Ccc}\label{equ:them_pres_apperr_one_assume_one}\\
		&\trule{\Cerr}{\Ctbot}{\Ctt}\nonumber
	\end{align}
	By inversion on equation~\eqref{equ:them_pres_apperr_one_assume_one} we also get
	$\trule{\hat{v}}{\Ctype_{\hat{v}}}{\Cff}$. As $\hat{v}\in\{n,\Cctorvnabbr\}$, we know that
	$\Fhasnfun{\Ctype_{\hat{v}}}$ holds and thus, $\Ccc=\Ctt$. But then, $\Ccc=\Ccc'$ and
	$\subtype{t'}{t}$.
	\item\case{$e=\Fapp{(\Frec{f}{x}{e})}{\Cerr} \hookrightarrow e'=\Cerr$}
	By assumption we have 
	\begin{align}
		&\trule{\Fapp{(\Frec{f}{x}{e})}{\Cerr}}{t}{\Ccc}\label{equ:them_pres_apperr_two_assume_one}\\
		&\trule{\Cerr}{\Ctbot}{\Ctt}\nonumber
	\end{align}
	By inversion on equation~\eqref{equ:them_pres_apperr_two_assume_one} we also get
	$\trule{\Cerr}{\Ctbot}{\Ctt}$. Thus, $\Ccc_2=\Ctt$ which implies $\Ccc=\Ctt$. But then, $\Ccc=\Ccc'$ and
	$\subtype{t'}{\Ftapp{\Ctype_1}{\Ctype_2}}$.
	\item\case{$e=\Fmatch{\Cerr}{\Cpatterns} \hookrightarrow e'=\Cerr$}
	by assumption we have 
	\begin{align}
		&\trule{\Fmatch{\Cerr}{\Cpatterns}}{t}{\Ccc}\label{equ:them_pres_matcherr_two_assume_one}\\
		&\trule{\Cerr}{\Ctbot}{\Ctt}\nonumber
	\end{align}
	by inversion on equation~\eqref{equ:them_pres_matcherr_two_assume_one} we also get
	$\trule{\Cerr}{\Ctbot}{\Ctt}$. Thus, $\Ccc=\Ctt$. But then, $\Ccc=\Ccc'$ and
	$\subtype{t'}{\Ctype_p}$.
	\item\case{$e=\Fmatch{\Cctorvnabbr}{\Cemptypattern} \hookrightarrow e'=\Cerr$}
	by assumption we have 
	\begin{align}
		&\trule{\Fmatch{\Cctorvnabbr}{\Cemptypattern}}{t}{\Ccc}\label{equ:them_pres_matchempty_assume_one}\\
		&\trule{\Cerr}{\Ctbot}{\Ctt}\nonumber
	\end{align}
	by inversion on equation~\eqref{equ:them_pres_matchempty_assume_one} we also get
	$\pruleenv{\Ctt;\Ctype_0}{\Cemptypattern}{\Ctype_p}{\Ccc_p}$. By \textsc{T-Pattern-Empty} we know that
	$\Ctype_p=\Ctbot$ and $\Ccc=\Ctt$. Thus, $t=\Ctbot$ and $\Ccc=\Ctt$. But then, $\Ccc=\Ccc'$ and
	$\subtype{t'}{t}$.
	\item\case{$e=C(v_1,\dots,v_n,\Cerr,e_1,\dots,e_m) \hookrightarrow e'=\Cerr$}
	Assumptions give us
	\begin{align}
		&\trule{C(v_1,\dots,v_n,\Cerr,e_1,\dots,e_m)}{t}{\Ccc}\label{equ:them_pres_ctorerr_assume_one}\\
		&\trule{\Cerr}{\Ctbot}{\Ctt}\nonumber
	\end{align}
	By inversion on equation~\eqref{equ:them_pres_ctorerr_assume_one} we get, amongst others,
	$\trule{\Cerr}{\Ctbot}{\Ctt}$, and thus, we can follow by \textsc{T-Constructor}, that $\Ccc=\Ctt$.
	We can now conclude $\subtype{t'}{t}$ and $\Ccc=\Ccc'$.
\end{description}
\end{proof}

\begin{proof}[\cref{thm:Failure}]
    For the simple cases $e=\{n,\Frec{f}{x}{e},\Cctorvnabbr,x\}$ the assumption
    does not hold, as $\trule{e}{\Ctype}{\Cff}$.

\begin{description}
    \item\case{$e=\Cerr$} Trivially holds.
    \item\case{$e=\Cctorenabbr$} Holds by inversion, application of the induction
    hypothesis and application of \textsc{SCtorErr}.
    \item\case{$e=\Fapp{e_1}{e_2}$}
    \begin{itemize}
        \item Wlog assume $\vDash \cT(\Ccc_1)$: Holds by inversion, application of the induction hypothesis
            and application of \textsc{SAppErr1}
        \item Wlog assume $\vDash \cT(\Ccc_2)$: Holds by inversion, application of the induction hypothesis
            and application of \textsc{SAppErr2}.
        \item Wlog assume $\vDash \cT(\Fhasnfun{\Ctype_1})$: We can assume that either $\cV(e_1)\!\Uparrow$ or
        $\cV(e_1)\hookrightarrow^* v$. For the former, clearly $\cV(e)\!\Uparrow$. For the latter,
        we apply \cref{thm:Preservation} and deduce that $v$ cannot be a function. Thus our claim holds
        by \textsc{SAppErr1}.
        \item Wlog assume $\vDash \cT(\Ftapp{\Ctype_1}{\Ctype_2})$, $\nvDash \cT(\Ccc_1)$,
        and $\nvDash \cT(\Ccc_2)$: We know that $\cV(e)\hookrightarrow^* \Fapp{(\Frec{f}{x}{e'})}{v}\stackrel{\textsc{SApp}}{\hookrightarrow} e''$. By inversion and application of the induction hypothesis
        our claim holds.
    \end{itemize}
    \item\case{$e=\Fmatch{e_0}{[\overline{\Cctorxnabbr \rightarrow e'}]}$}
    \begin{itemize}
        \item Wlog assume $\vDash \cT(\Ccc_0)$: By inversion, application of the induction
        hypothesis, and \textsc{SMatchErr} our claim holds.
        \item Wlog assume $\vDash \cT(\Ccc_p)$: By inversion we get
        \[\pruleenv{\Ctt;\Ctype_0}{\Cpatterns}{\Ctype_p}{\Ccc_p}\]
        Again by inversion we know that $\cV(\Cexp_0)\hookrightarrow^* \Cctorvnabbr$ with
        $\trule{\Cctorvnabbr}{\Ctctortnabbr}{\Cff}$. Thus we can apply the induction hypothesis
        and get
        \begin{itemize}
            \item that there is no matching constructor. Thus \[\cV(\Cexp)\stackrel{\textsc{SMatchErr}}{\hookrightarrow^*}\cV(\Fmatch{\Cctorvnabbr}{\Cemptypattern}\] and by \textsc{SMatchNextErr} our claim holds.
            \item that there is a matching constructor. Thus
                \[\cV(\Cexp)\stackrel{\textsc{SMatchNext}}{\hookrightarrow^*}\cV(e'')\hookrightarrow\cV(e')\hookrightarrow^*\Cerr\]
        \end{itemize}
    \end{itemize}
\end{description}

For the pattern-matching cases we have get the following:
\begin{description}
    \item\case{TPatternEmpty} Trivial follows that there is not matching constructor.
    \item\case{TPatternNext} There are two cases:
    \begin{itemize}
        \item If the first constructor matches the constructor value, then $\nvDash \cT(\Ccc_0)$ because
        it is the first match and \cref{lem:UnsatisfiabilityAfterMatchingPatterns}.
        Furthermore, we have $\vDash \cT(\Fhasctor{C}{\Ctype_0})$. By assumption we conclude
        that $\vDash\cT(\Ccc_e)$ and thus apply the induction hypothesis on
        \[\truleenv{x_i: \Ftget{\Ctype_0}{\Cctor}{i}}{\Cexp}{\Ctype_{\Cexp}}{\Ccc_{\Cexp}}\]
        and our claim holds.
        \item If the first constructor does not match, we apply inversion and the induction
        hypothesis and on the remaining patterns and get:
        \begin{itemize}
            \item if there is no matching constructor our claim immediately holds.
            \item if there is a matching constructor $\Cctor_j$ with body expression $\Cexp_j$
            and $\cV(\Cexp_j)\hookrightarrow^* \Cerr$ or $\cV(\Cexp_j)\!\Uparrow$, then our
            claim holds for $i=j+1$.
        \end{itemize}
    \end{itemize}
\end{description}

\end{proof}

}{}

\end{document}

%% file: commonmacros.tex
\renewcommand{\case}[1]{\emph{Case} \textsc{#1}:}
\newcommand{\subcase}[1]{\emph{Subcase} \textsc{#1}:}

\newcommand{\comment}[1]{}

\newcommand{\Fdom}[1]{\ensuremath{\mathbf{dom}(#1)}}

\newcommand{\newmcommand}[3][\empty]{
  \ifthenelse{\equal{#1}{\empty}}
    {\newcommand{#2}{\ensuremath{#3}}}
    {\newcommand{#1}[#2]{\ensuremath{#3}}}
}

\newcommand{\newXcommand}[3][\empty]{
  \ifthenelse{\equal{#1}{\empty}}
    {\newcommand{#2}{\ensuremath{#3}\xspace}}
    {\newcommand{#1}[#2]{\ensuremath{#3}\xspace}}
}

\newcommand{\cC}{\mathcal{C}}

\newcommand{\cE}{\mathcal{E}}

\newcommand{\cJ}{\mathcal{J}}

\newcommand{\cT}{\mathcal{T}}

\newcommand{\cV}{\mathcal{V}}


%% file: localmacros.tex
\newcommand{\Clangname}{\ensuremath{\lambda_C}\xspace}

\newcommand{\OnlineLocation}{\url{http://www.informatik.uni-freiburg.de/~jakobro/stpa/}}

\newXcommand{\Cctor}{C}
\newXcommand{\Cctord}{D}
\newXcommand{\Cctoren}{\Cctor(\Cexp_1,\dots,\Cexp_n)}
\newXcommand{\Cctorenabbr}{\Cctor(\overline{\Cexp})}
\newXcommand{\Cctorvn}{\Cctor(\Cval_1,\dots,\Cval_n)}
\newXcommand{\Cctorvnabbr}{\Cctor(\overline{\Cval})}
\newXcommand{\Cctorxn}{\Cctor(x_1,\dots,x_n)}
\newXcommand{\Cctorxnabbr}{\Cctor(\overline{x})}
\newXcommand{\Cctordxmabbr}{\Cctord(\overline{x})}
\newXcommand{\Cctors}{\cC}
\newXcommand{\Cenv}{\Gamma}
\newXcommand{\Cerr}{\mathrm{err}}
\newmcommand{\Cevaluatesto}{\longrightarrow}
\newXcommand{\Cexp}{e}
\newXcommand{\Chole}{\square}
\newmcommand{\Cmatch}{\mathbf{match}}
\newXcommand{\Cinter}{\cJ}
\newXcommand{\Cinterp}{{\cJ'}}
\newXcommand{\Cpatterns}{P}
\newmcommand{\Crec}{\mathbf{rec}}
\newXcommand{\Ctype}{\tau}
\newXcommand{\Ctypes}{\cT}
\newXcommand{\Ctvar}{\alpha}
\newXcommand{\Ctint}{\mathrm{int}}
\newXcommand{\Ctbot}{\bot}
\newXcommand{\Ccc}{\phi}
\newXcommand{\Ctt}{\mathtt{tt}}
\newXcommand{\Cff}{\mathtt{ff}}
\newXcommand{\Ctctortn}{\Cctor(\Ctype_1,\dots,\Ctype_n)}
\newXcommand{\Ctctortnabbr}{\Cctor(\overline{\Ctype})}
\newXcommand{\Cunfold}{\mathrm{unfold}}
\newXcommand{\Cval}{v}
\newXcommand{\CvalNoFun}{\hat{v}}
\newmcommand{\Cwith}{\mathbf{with}}
\newXcommand{\Cemptypattern}{[\ ]}

\newcommand{\Ftinter}[1]{\llbracket #1 \rrbracket_\Cinter}
\newcommand{\Ftinterp}[1]{\llbracket #1 \rrbracket_\Cinterp}
\newcommand{\Fapp}[2]{(#1~#2)}
\newcommand{\Fevals}[2]{#1\ \Cevaluatesto\ #2}
\newcommand{\Fmatch}[2]{\Cmatch~#1~\Cwith~#2}
\newcommand{\Frec}[3]{\Crec~#1~#2=#3}
\newcommand{\Fsemanticrule}[3]{\textsc{#1}\ #2 \Cevaluatesto #3}
\newcommand{\Ftfun}[3]{\ensuremath{\mu X.\forall #1 \left[#3\right]. #2}}
\newcommand{\Ftapp}[2]{(#1 \mathbin{@_\Ctype} #2)}
\newcommand{\Ftget}[3]{#1\!\downarrow^{#2}_{#3}}
\newcommand{\Fhasctor}[2]{#1 \in #2}
\newcommand{\Fhasnctor}[2]{#1 \notin #2}
\newcommand{\Fhasnfun}[1]{\forall \notin #1}
\newcommand{\Fcc}[2]{(#1 \mathbin{@_\Ccc} #2)}

\newcommand{\auxcases}{[\Cctorxnabbr \rightarrow \Cexp]}
\newcommand{\auxsubst}[2]{#1[#2]}
\newcommand{\ectordxbar}{\Cctord(\bar{x})}
\newcommand{\phasdtor}[1]{D \in #1}
\newcommand{\phasndtor}[1]{D \notin #1}

\newcommand{\sep}{\mid}

\newcommand\append{\ensuremath{\mathbin{+\mkern-10mu+}}}

\newcommand{\typecrashsep}{\&}
\newcommand{\trule}[3]{\Cenv\vdash #1 : #2\ \typecrashsep \ #3}
\newcommand{\truleenv}[4]{\Cenv,#1\vdash #2 : #3\ \typecrashsep \ #4}
\newcommand{\truleenvown}[4]{#1\vdash #2 : #3\ \typecrashsep \ #4}
\newcommand{\pruleenv}[4]{#1;\Cenv\vdash_p #2 : #3\ \typecrashsep \ #4}
\newcommand{\pruleenvg}[5]{#1;\Cenv,#2\vdash_p #3 : #4\ \typecrashsep \ #5}
\newcommand{\pruleenvown}[5]{#1;#2\vdash_p #3 : #4\ \typecrashsep \ #5}

\newcommand{\truleinteger}{\inferrule[T-Integer]{\ }
 {\trule{n}{\Ctint}{\Cff}}
 }

\newcommand{\trulerec}{%
\inferrule[T-Rec]{
	\truleenv{x_r:\Ctvar_r,f_r:X}
		{\Cexp}{\Ctype_{\Cexp}}{\Ccc_{\Cexp}}\\
		\Ctvar_r\ \textnormal{fresh}\quad x_r,f_r\notin\Fdom{\Cenv}}
 {\trule{\Frec{f_r}{x_r}{\Cexp}}{\Ftfun{\Ctvar_r}{\Ctype_{\Cexp}}{\Ccc_{\Cexp}}}{\Cff}}
 }

\newcommand{\truleapp}{\inferrule[T-FunApp]{
	\trule{\Cexp_1}{\Ctype_1}{\Ccc_1} \\\\ \trule{\Cexp_2}{\Ctype_2}{\Ccc_2}}
 {\trule{\Fapp{\Cexp_1}{\Cexp_2}}{\Ftapp{\Ctype_1}{\Ctype_2}}{\Fcc{\Ctype_1}{\Ctype_2}
 	\vee \Ccc_1 \vee \Ccc_2 \vee \Fhasnfun{\Ctype_1}}}
 }

\newcommand{\truleid}{\inferrule[T-Identifier]{\Cenv(x) = \Ctype}
 {\trule{x}{\Ctype}{\Cff}}
 }

\newcommand{\trulepm}{\inferrule[T-Pattern-Matching]
{\pruleenv{\Ctt;\Ctype_0}{\Cpatterns}{\Ctype_p}{\Ccc_p} \\\\
	\trule{\Cexp_0}{\Ctype_0}{\Ccc_0}}
  {\trule{\Fmatch{\Cexp_0}{\Cpatterns}}{\Ctype_p}{\Ccc_0 \vee \Ccc_p}}
  }

\newcommand{\trulepmempty}{\inferrule[T-Pattern-Empty]
 {\ }
 {\pruleenv{\Ccc_0;\Ctype_0}{\Cemptypattern}{\Ctbot}{\Ccc_0}}
 }

\newcommand{\trulepmnext}{\inferrule[T-Pattern-Next]{%
  \pruleenv{\Ccc_0\wedge ((\Fhasctor{\Cctor}{\Ctype_0} \wedge \Ccc_e) \vee \Fhasnctor{\Cctor}{\Ctype_0});\Ctype_0}
  	{\Cpatterns}{\Ctype'}{\Ccc'}
  \\\\
  \truleenv{x_i: \Ftget{\Ctype_0}{\Cctor}{i}}{\Cexp}{\Ctype_{\Cexp}}{\Ccc_{\Cexp}}\quad i=1,\dots,n}
  {\pruleenv{\Ccc_0;\Ctype_0}{[\Cctorxn\rightarrow \Cexp] \append \Cpatterns}{\Ctype' \cup \Ctype_\Cexp}{\Ccc'}}
}

\newcommand{\trulector}{\inferrule[T-Constructor]{
	\forall i \in \{1,\dots,n\}: \trule{\Cexp_i}{\Ctype_i}{\Ccc_i}}
 {\trule{\Cctoren}{\Ctctortn}{\bigvee \overline{\Ccc}}}
 }

\newcommand{\truleerr}{\inferrule[T-Error]{\ }
  {\trule{\Cerr}{\Ctbot}{\Ctt}}
}

\newcommand{\subtype}[2]{#1 \leq #2}

\newcommand{\trulesubgeneral}{\inferrule[T-Sub]
  {\trule{\Cexp}{\Ctype}{\Ccc}\\ \subtype{\Ctype}{\Ctype'} \\ \Ccc'\rightarrow \Ccc}
  {\trule{\Cexp}{\Ctype'}{\Ccc'}}
}

\newcommand{\trulesubbot}{\inferrule[S-Bot]{\ }
  {\subtype{\Ctbot}{\Ctype}}
}

\newcommand{\trulesubunion}{\inferrule[S-Union]{\ }
  {\subtype{\Ctype}{\Ctype\cup \Ctype'}}
}

\newcommand{\trulesubrefl}{\inferrule[S-Refl]{\ }
  {\subtype{\Ctype}{\Ctype}}
}

\newcommand{\trulesubctor}{\inferrule[S-Ctor]
  {\subtype{\bar{\Ctype}}{\bar{\Ctype'}}}
  {\subtype{\Cctor(\bar{\Ctype})}{\Cctor(\bar{\Ctype'})}}
}

\newcommand{\trulesubfun}{\inferrule[S-Fun]
  {\subtype{\Ctype}{\Ctype'}\\ \Ccc'\rightarrow \Ccc}
  {\subtype{\Ftfun{\Ctvar}{\Ctype}{\Ccc}}{\Ftfun{\Ctvar}{\Ctype'}{\Ccc'}}}
}

%% file: abstract.tex
\begin{abstract}

Dynamic languages are praised for their flexibility and
expressiveness, but static analysis often yields many
false positives and verification is cumbersome for lack of structure.
Hence, unit testing is the prevalent incomplete method for
validating programs in such languages. 

Falsification is an alternative approach that uncovers definite errors
in programs. A falsifier computes a set of inputs that definitely
crash a program.

Success typing is a type-based approach to document programs in
dynamic languages. We demonstrate that success typing is, in fact, an
instance of falsification by mapping success (input) types into
suitable logic formulae. Output types are represented by recursive types.
We prove the correctness of our mapping (which establishes that
success typing is falsification) and we report some experiences with
a prototype implementation.

\end{abstract}